\documentclass[reprint,amsmath,amssymb,aps,prc
superscriptaddress,
groupedaddress,
unsortedaddress,
runinaddress,
frontmatterverbose,
]{revtex4-1}
\usepackage{graphicx}
\usepackage{dcolumn}
\usepackage{bm}
\begin{document}
\title{Swift Heavy Ion Irradiation Induced Modifications in Structural, Microstructural, Electrical and Magnetic Properties of Mn Doped $SnO_{2}$ Thin Films}
\author{Sushant Gupta$^1$$^*$, Fouran Singh$^2$ and B. Das$^1$}
\affiliation{$^1$Department of Physics, University of Lucknow, \\Lucknow-226007, India\\
$^2$Inter University Accelerator Centre (IUAC), \\New Delhi-110067, India\\\\
$^*$E-mail: sushant1586@gmail.com}
\begin{abstract}
In this paper, we have presented the impact of swift heavy ion beam irradiation on the structural, microstructural, electrical and magnetic properties of $Sn_{0.9}Mn_{0.1}O_{2}$ thin films. The structural and electrical results have been interpreted by using properties of native or point defects, whereas the magnetic and morphological variations have been explained in terms of conductivity of material. Efforts have been made to summarize the properties of all possible charged and neutral point defects ($V_{Sn}^{4-}$, $Sn_{i}^{4+}$, $V_{O}^{0}$, $O_{i}^{2-}$, $Sn_{O}^{4+}$, $O_{Sn}^{2-}$, $H_{O}^{+}$) and afterwards from the correlation between experimentally-observed and theoretically-calculated results various interesting conclusions have been drawn.
\end{abstract}
\maketitle
\section{Introduction}
Transparent conducting oxides (TCO) are a unique class of material that combines two features, electrical conductivity and optical transparency, which are not typically found in the same material [1-4]. A good TCO material should have a carrier concentration of the order of $10^{20}$ $cm^{-3}$, a resistivity of the order of $10^{-3}$ $\Omega$ cm, and an optical transmission above 80\% in the visible range of the electromagnetic spectra. Tin oxide ($SnO_{2}$) thin films with such unique features have been extensively used in optoelectronic applications for flat panel displays, smart windows, solar cells, and electromagnetic interference shielding windows [5, 6]. Recently there is an increased interest to introduce magnetic functionality in tin oxide semiconductors due to their promising applications in spintronics [7-12]. The tin oxide semiconductor can be made ferromagnetic by doping with transition-metal (TM) ions. The first report of high Curie temperature ferromagnetism in tin oxide thin films was by Ogale et. al. [13], who reported a Curie temperature $T_{c}$ = 650 K in pulsed laser deposited rutile $(Sn_{1-x}Co_{x})O_{2}$ thin films with x = 5-27\%, and an amazingly giant magnetic moment of (7.5$\pm$0.5)$\mu_{B}$ per Co ion. High Curie temperature ferromagnetism was latterly reported for $(Sn_{1-x}Ni_{x})O_{2}$ with x = 5\% [14, 15], $(Sn_{1-x}V_{x})O_{2}$ with x = 7\% [16], $(Sn_{1-x}Cr_{x})O_{2}$ with x = 5\% [17], and $(Sn_{1-x}Fe_{x})O_{2}$ with x = 14\% [18] \& x = 0.5-5\% [19]. Gopinadhan et. al. [20] investigated  $(Sn_{1-x}Mn_{x})O_{2}$ (with x = 10\%) thin films deposited by spray pyrolysis method and found ferromagnetic behavior above room temperature with low magnetic moment of 0.18$\pm$0.04 $\mu_{B}$ per Mn ion. Fitzgerald et. al. [21] studied 5\% Mn-doped $SnO_{2}$ bulk ceramic samples and reported a Curie temperature of $T_{c}$ = 340 K with magnetic moment of 0.11 $\mu_{B}$ per ordered Mn ion. On the contrary, Duan et. al. [22] reported an antiferromagnetic superexchange interaction in Mn-doped $SnO_{2}$ nanocrystalline powders and Kimura et. al. [23] observed paramagnetic behavior of Mn-doped $SnO_{2}$ thin films. Apart from this some other experiments were also carried out by various research groups on $SnO_{2}$ based dilute magnetic semiconductors (DMS) and reported interesting results regarding the absence or presence of ferromagnetism [24-35]. DMS based on $SnO_{2}$ could be useful for a variety of applications requiring combined optical and magnetic functionality [36]. Several devices such as spin transistors, spin light-emitting diodes, very high-density nonvolatile semiconductor memory, and optical emitters with polarized output have been proposed using $Sn_{1-x}(TM)_{x}O_{2}$ materials [37-40].

The origin of n-type conductivity in tin oxide semiconductor is a subject of controversy [41-43]. Thin films of pure, but non-stoichiometric, $SnO_{2}$ mostly show free-electron concentrations in the $n \sim 10^{20}$ $cm^{-3}$ range. The prevalence of such high electron concentrations has historically been explained by oxygen vacancies ($V_{O}$) and tin interstitials ($Sn_{i}$) [44]. Recent density functional theory (DFT) calculations and experiments [41, 42], however, have provided evidence that usual suspects such as $V_{O}$ and $Sn_{i}$ are actually not responsible for n-type conductivity in majority of the
cases. These calculations indicate that the oxygen vacancies are a deep donor, whereas tin interstitials are too mobile to be stable at room temperature [42]. As discussed further in this paper, hydrogen is an ubiquitous impurity that can lead to n-type doping [41, 42, 45-50].

Swift heavy ion (SHI) irradiation, in which an energetic ion beam is allowed to pass through a material, is a very effective method to induce structural/microstructural modifications in materials and has been used to tailor the properties of various materials including insulators, metals, semiconductors, and polymers [51-57]. High electronic energy released by the SHI beams in a very short span of time produces significant excitation in the crystal lattice, causing changes in the structural, optical, electrical and magnetic behaviour of the materials. The SHI irradiation is known to generate controlled defects (points/clusters and columnar) and structural disorder [53, 55-57]. The parameters of SHI beam such as mass and energy of the ion and the properties of the target materials such as conductivity and microstructure, play crucial role in defect generation and engineering [52, 53, 55-57]. The ion energy and its mass, decide the magnitudes of the electronic as well as nuclear energy losses. The other ion beam parameter, fluence, dictates the number of defects (created). The number of ion tracks per $cm^{2}$ is the same as that of fluence (ions/$cm^{2}$) because each ion creates an ion-track [53]. Modifications produced by SHI are quite different from that of low energy ions. There is a threshold of electronic energy loss, beyond which the creation of columnar defect or the latent track occurs in the materials [51, 52, 58]. This threshold energy depends on the electron-phonon coupling (g) and conductivity of the target material [51, 59-61]. It can be up to about a few hundred $eV/\dot{A}$ for polymers and other insulators [61, 62] and it can be a few $keV/\dot{A}$ for metals [63-65]. There are certain materials like Ag, Cu, Ge, Si etc., in which track formation is not possible at any energy with monoatomic ion beams [66-69]. Investigations of defects produced by swift heavy ion irradiation in oxide semiconductors such as $SnO_{2}$ and $In_{2}O_{3}$ have become an important area of research in view of multifunctional magneto-optoelectronic devices. It would be interesting to carry out a detailed study on modifications induced by 120 MeV gold ions on conducting tin oxide $(n \geq 10^{20} cm^{-3})$ thin films.
\section{Experimental details}
Thin films of $SnO_{2}$ and $Sn_{0.9}Mn_{0.1}O_{2}$ were deposited by spray pyrolyzing a mixture of aqueous solutions of tin (II) chloride di-hydrate ($SnCl_{2}.$ $2H_{2}O$) and manganese (II) acetate tetra-hydrate $((CH_{3}COO)_{2}Mn.4H_{2}O)$ on glass/quartz substrates at $450^{o}C$. An amount of 11.281 gm of $SnCl_{2}.2H_{2}O$ (Sigma Aldrich purity $>$ 99.99\%) was dissolved in 5 ml of concentrated hydrochloric acid by heating at $90^{o}C$ for 15 min. The addition of hydrochloric acid rendered the solution transparent, mostly, due to the breakdown of the intermediate polymer molecules. The transparent solution subsequently diluted with ethyl alcohol formed the precursor. To achieve Mn doping, $(CH_{3}COO)_{2}Mn.4H_{2}O$ was dissolved in ethyl alcohol and added to the precursor solution. The amount of $(CH_{3}COO)_{2}Mn.4H_{2}O$ to be added depends on the desired doping concentration. The overall amount of spray solution in each case was made together 50 ml. The repeated experiments of each deposition showed that the films could be reproduced easily. Pyrex glass and fused quartz slides (10 mm $\times$ 10 mm $\times$ 1.1 mm), cleaned with organic solvents, were used as substrates for various studies. During deposition, the solution flow rate was maintained at 0.2 ml/min by the nebulizer (droplet size 0.5-10 $\mu$m). The distance between the spray nozzle and the substrate as well as the spray time was maintained at 3.0 cm and 15 min, respectively. One set of as-deposited $SnO_{2}$ films were annealed in air for 4 h at $850^{o}C$.

The as-deposited $Sn_{0.9}Mn_{0.1}O_{2}$ films were irradiated with 120 MeV $Au^{9+}$ ions at six different fluences ($1\times10^{11}$, $3\times10^{11}$, $1\times10^{12}$, $3\times10^{12}$, $1\times10^{13}$ and $3\times10^{13}$ ions/$cm^{2}$) in a 15 MV pelletron accelerator at Inter University Accelerator Centre (IUAC), New Delhi. During irradiation the samples were attached to a massive sample holder using a double sided tape. The irradiation was performed in the direction nearly perpendicular to the sample surface. Ion beam was focused to a spot of 1 mm diameter and then scanned over 1 $cm^{2}$ area using a magnetic scanner. The vacuum of $10^{-6}$ Torr was maintained during irradiation experiments. During irradiation the current of the ion beam was kept at 0.5 PnA. The fluence values were measured by collecting the charges falling on the sample mounted on an electrically insulated sample holder placed in secondary electron suppressed geometry. Ladder current was integrated with a digital current integrator and the charged pulses were counted using scalar counter.

The gross structure and phase purity of all films were examined by X-ray diffraction (XRD) technique using a Bruker AXS, Germany X-ray diffractometer (Model D8 Advanced) operated at 40 kV and 60 mA. In the present study, XRD data of unirradiated and irradiated thin films were recorded in the scanning angle ($2\theta$) range $20^{o}- 60^{o}$ using $Cu-K_{\alpha}$ radiations ($\lambda$ = 1.5405 ${\AA}$). All the diffraction patterns were collected under a slow scan with a $0.01^{o}$ step size and a counting velocity of $0.3^{o}$ per minute. The experimental peak positions were compared with the data from the database Joint Committee on Powder Diffraction Standards (JCPDS) and Miller indices were assigned to these peaks. Transmission Electron Microscopy (TEM) measurements were carried out on a Tecnai $20^{2}$G microscope with an accelerating voltage of 200 kV. All the images were digitally recorded with a slow scan charge-coupled device camera (image size 688 $\times$ 516 pixels), and image processing was carried out using the digital micrograph software. The TEM data were used for the study of grain size distribution and the crystalline character of the prepared samples. These TEM micrographs were also used to identify secondary phases present, if any, in the $Sn_{0.9}Mn_{0.1}O_{2}$ matrix. Surface morphology of the samples was analyzed by means of Scanning Electron Microscopy (SEM) using a MIRA II LMH, TESCAN: field emission scanning electron microscope with a maximum resolution of 1.5 nm at 30 kV. Atomic Force Microscopy (AFM) was performed with Multi Mode SPM (Digital Instrument Nanoscope E) in AFM mode to examine the microstructural evolution and root mean square (rms) surface roughness of the sample before and after irradiation. Hall measurements were conducted at room temperature to estimate the film resistivity ($\rho$), donor concentration (n) and carrier mobility ($\mu$) by using the van der Pauw geometry employing Keithley's Hall effect card and switching the main frame system. A specially designed Hall probe on a printed circuit board (PCB) was used to fix the sample of the size 10 mm $\times$ 10 mm. Silver paste was employed at the four contacts. The electrical resistivity and the sheet resistance of the films were also determined using the four-point probe method with spring-loaded and equally spaced pins. The probe was connected to a Keithley-voltmeter-constant-current source system and direct current and voltage were measured by slightly touching the tips of the probe on the surface of the films. Multiple reading of current and the corresponding voltage were recorded in order to get average values. Optical absorption measurement was performed at room temperature within a wavelength range of 200-800 nm using a Cary 5000 UV-Vis spectrophotometer having spectral resolution of 0.05 nm in the UV-Vis range. As a reference, 100\% baseline signals were displayed before each measurement. Magnetic measurements were carried out as a function of temperature (5 to 300 K) and magnetic field (0 to $\pm$2 T) using a `EverCool 7 Tesla' SQUID magnetometer. Measurements were carried out on small size films placed in a clear plastic drinking straw. The data reported here were corrected for the background signal from the sample holder (clear plastic drinking straw) independent of magnetic field and temperature. Thickness of the deposited films was estimated by an Ambios surface profilometer and was approximately 500 nm.
\begin{table*}[htbp]
\renewcommand{\arraystretch}{1.5}
\caption{The values of $S_{n}$, $S_{e}$, $R_{p}$ for 120 MeV $Au^{9+}$-ion-irradiated $Sn_{0.9}Mn_{0.1}O_{2}$ thin film deposited on a glass substrate estimated by SRIM [70, 71]. The value of $S_{eth}$ and possibility of track formation are also mentioned.}
\vspace{5mm}
\centering
\begin{tabular}{||c|c|c||}
\hline
\hline
Parameters & $Sn_{0.9}Mn_{0.1}O_{2}$ & Glass substrate\\
\hline
\hline
Nuclear energy loss & 0.485 & 0.216 \\
  ($S_{n}$ in keV $nm^{-1}$) &  &  \\
  \hline
  Electronic energy loss & 27.79 & 14.88 \\
  ($S_{e}$ in keV $nm^{-1}$) &  &  \\
  \hline
  Threshold electronic energy & 31.92 & 3.033 \\
  loss ($S_{eth}$ in keV $nm^{-1}$)&  &  \\
  \hline
  Range ($\mu$m) & 8.33 & 15.28 \\
  \hline
  Track formation & Not possible & Possible \\
\hline
\hline
\end{tabular}
\end{table*}
\section{Results and discussion}
\begin{figure}
  \centering
  \includegraphics[height=7.0cm, width=8.5cm]{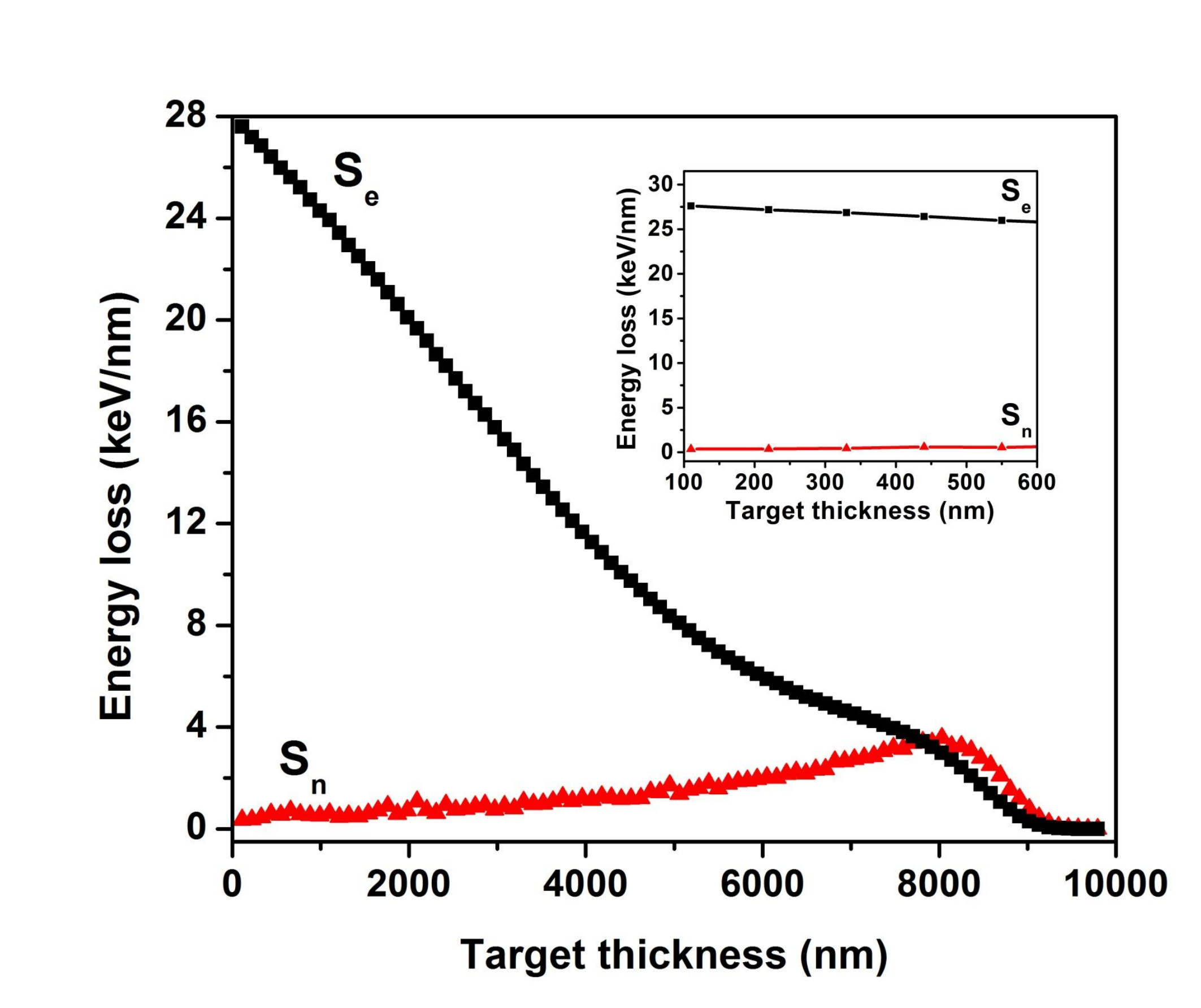}\\
  \caption{Variation of $S_{e}$ and $S_{n}$ of 120 MeV $Au^{9+}$ ions incident on $Sn_{0.9}Mn_{0.1}O_{2}$ thin film as a function of film thickness. The inset shows the almost constant value of $S_{e}$ for even 500 nm depth from the surface of the thin film.}\label{1}
\end{figure}
The electronic energy loss $S_{e}$, nuclear energy loss $S_{n}$ and range $R_{p}$ of 120 MeV $Au^{9+}$ ions in the $Sn_{0.9}Mn_{0.1}O_{2}$ thin film are 27.79 $keV/nm$, 0.485 $keV/nm$ and 8.33 $\mu m$, respectively. These values were calculated through the standard simulation program SRIM [70, 71]. From the calculated values, it is clearly evident that the $S_{e}$ is about two orders of magnitude more than that of $S_{n}$ and the range is much larger than the film's thickness ($\sim$ 500 nm). The variation of energy loss (both electronic and nuclear) for 120 MeV gold ions incident on $Sn_{0.9}Mn_{0.1}O_{2}$ thin film with the film thickness from the surface is shown in Fig. 1. Near the surface of the film, $S_{e}$ exceeds $S_{n}$ by two orders of magnitude and is almost constant throughout the film thickness, as shown in the inset of Fig. 1. This reveals that the morphological and structural changes in $Sn_{0.9}Mn_{0.1}O_{2}$ thin film on irradiation by 120 MeV $Au^{9+}$ ions are due to electronic excitations. The values of $S_{n}$, $S_{e}$ and $R_{p}$ for 120 MeV $Au^{9+}$-ions-irradiated $Sn_{0.9}Mn_{0.1}O_{2}$ thin film (thickness $\sim$ 500 nm) on glass substrate (thickness $\sim$ 1.1 mm) are mentioned in the Table I along with the threshold value of electronic energy loss ($S_{eth}$), calculated later in the paper.
\subsection{Threshold energy loss}
To create latent tracks or to induce amorphization in crystalline materials a certain threshold value of $S_{e}$ is required and according to Szenes' [61] ``thermal spike model'' this threshold value $S_{eth}$ depends on the material parameters such as electron-phonon coupling efficiency g, density $\rho$, melting temperature $T_{m}$, average specific heat C, irradiation temperature $T_{irr}$, and initial width of the thermal spike a(0). The Szenes' equation for the threshold energy can be expressed as follows:
\begin{equation}\label{1}
  S_{eth} = \frac{\pi\rho C(T_{m} - T_{irr})a^{2}(0)}{g}
\end{equation}
The value of a(0), for semiconductors, depends on their energy band gap $E_{g}$, while for insulators it is almost constant. This ion induced thermal spike width a(0) in semiconductors can be best approximated by the expression $a(0) = b + c (E_{g})^{\frac{-1}{2}}$ where b and c are constants. Using an $E_{g}$ value of 4.3 eV (see Section III.H) for $Sn_{0.9}Mn_{0.1}O_{2}$, the value of a(0) was determined from the plot of $(E_{g})^{\frac{-1}{2}}$ versus a(0) as given in reference-59, and was found to be $\sim 6.2$ nm. The electron-phonon coupling efficiency (g) can be defined as the ability of electrons to transfer their energy to the lattice. According to Szenes [59, 61], the value of g depends on electron concentration (n). It is found that, for conductive materials (n $\geq$ $10^{20}$ $cm^{-3}$), the value of g is $\sim$ 0.092 [59], while for insulators it is $\sim$ 0.4 [72]. In a later section-III.F, we have reported the results of electrical measurements. From these measurements, we can say as-deposited $Sn_{0.9}Mn_{0.1}O_{2}$ samples are in conducting state with n = $1.93 \times 10^{20}$ $cm^{-3}$. By substituting the values of $\rho$ = 6.99 $g/cm^{3}$, $T_{m}$ = 1898 K, $T_{irr}$ = 300 K, and C  = 0.349 J/(gK) [73, 74] in equation-1, one can obtain $S_{eth}$ = 31.92 $keV/nm$ for $Sn_{0.9}Mn_{0.1}O_{2}$. For the present case of 120 MeV $Au^{9+}$ ions irradiating the $Sn_{0.9}Mn_{0.1}O_{2}$ film, $S_{e}$ is 27.79 keV/nm (see Table-I), which is less than the threshold value ($S_{eth}$ = 31.92 $keV/nm$) required to produce tracks/melted zones. Therefore we expect that only point defects or clusters of point defects will be produced in $Sn_{0.9}Mn_{0.1}O_{2}$ film after irradiation. On the other hand, for glass substrate, the value of $S_{eth}$ is 3.033 keV/nm (calculated by using $\rho$ = 2.53 $g/cm^{3}$, a(0) = 4.5 nm, C = 0.88 J/(gK), $T_{m}$ = 1673 K, $T_{irr}$ = 300 K, g = 0.4), which is much smaller than the available $S_{e}$ of 14.88 keV/nm induced by 120 MeV $Au^{9+}$ ions. Hence, track formation is only possible in the glass substrate but not in the $Sn_{0.9}Mn_{0.1}O_{2}$ film. This discontinuity in track formation can induce stress at the film-substrate interface.

It is important to understand that the way in which the huge energy ($S_{e}$ $\sim$ 27.79 keV/nm) is deposited to the system will modify the materials. The swift heavy ion initially interacts with the atomic electrons (electronic subsystem) of the target material and transfers its energy to electrons (valence and core electrons) in a time less than $10^{-16}$ s and the timescale of electron-ion interaction is very small, initially the lattice (atomic subsystem) of target material does not respond in that timescale [51, 52]. The transfer of the excess heat energy of the excited electronic subsystem to the atomic subsystem may lead to an increase in the local temperature of the target material and therefore a high energy region is developed in the close vicinity of the ion path. The amount of energy transferred (locally) depends on the coupling between the electronic and atomic subsystems. This coupling is known as electron-phonon coupling (g) and it can be defined as the ability of electrons to transfer their energy to the lattice. According to Szenes [59, 72], the value of g depends on the carrier concentration (n) in the target material. It is found that, for conductive materials ($n \geq 10^{20} cm^{-3}$), the value of g is $\sim$ 0.092 [59], while for insulators it is $\sim$ 0.4 [72]. The amorphized latent tracks are created only above a certain threshold value of $S_{e}$, which inversely depends on the electron-phonon coupling efficiency (g). Therefore, conductive targets will require very high beam energy to create latent tracks/melted zones. But if $S_{e} < S_{eth}$, then only point defects or cluster of point defects will create in target material after irradiation. A point to note here is that in insulator if  $S_{e} < S_{eth}$ then point defects will create locally around the ion path whereas in conductive material these defects will create uniformly through out the material. In this way, we can conjecture that the SHI irradiation can create tracks or point defects depending upon the initial state of the material. But if the system already contains such defects, then they can be annealed through the irradiation.
\subsection{Equilibrium substrate/film temperature}
\begin{table*}[htbp]
\renewcommand{\arraystretch}{1.4}
\caption{Summary of the irradiation conditions: beam fluence, beam current, irradiation period, and sample temperature in irradiation period.}
\vspace{5mm}
\centering
\begin{tabular}{||c|c|c|c||}
\hline
\hline
Beam fluence & Beam current & Irradiation period & Sample temperature\\
($ions/cm^{2}$) & (nA) & (sec)& in irradiation period (K)\\
\hline
\hline
$1 \times 10^{11}$ & 4.5 & 32 & Low temperature  \\
                         &  & & transient condition  \\
\hline
$3 \times 10^{11}$ & 4.5 & 96 & Low temperature \\
                         &  & &   transient condition\\
\hline
$1 \times 10^{12}$ & 4.5 & 320 & Low temperature\\
                               &  & & transient condition  \\
\hline
$3 \times 10^{12}$ & 4.5 & 960 & $\sim$ 834 K \\
                               &  & &   (not entire time)\\
\hline
$1 \times 10^{13}$ & 4.5 & 3200 & $\sim$ 834 K\\
                                &  & & (almost entire time)  \\
\hline
$3 \times 10^{13}$& 4.5 & 9600 & $\sim$ 834 K\\
                               &  & & (almost entire time)  \\
\hline
\hline
\end{tabular}
\end{table*}
The input power density can be calculated by using the following equation:-
$$P = \frac{(Irradiation~Energy \times Beam~Current)}{Scanning~Area}$$
In the present case the maximum input power density is
$$ P = \frac{(120~MeV \times 4.5~nA)}{1~cm^{2}} = 0.54 \frac{W}{cm^{2}}$$
The power of the ion beam deposited in the target can be removed mainly by thermal radiation and directed to the heat sink constituted by the vacuum chamber. Although the actual sample holder structures including the target chamber are complicated, but we have used the simplified Stefan's equation [75, 76] for temperature determination,
\begin{equation}\label{2}
P = \epsilon \sigma (T^{4} - T_{s}^{4}),
\end{equation}
where P is the input power density of high energy ion beam, $\sigma$ is the Stefan-Boltzmann constant of 5.67 $\times$ $10^{-12}$ $\frac{W}{cm^{2}K^{4}}$, $\epsilon$ is the effective emittance of the glass substrate, $T_{s}$ is the surrounding target chamber temperature ($\sim$ 300 K), and T is the equilibrium substrate/film temperature. According to Kumar et. al. [77], the value of effective emittance ($\epsilon$) for glass substrate is 0.2. By using the values of input power density (P = 0.54 $W/cm^{2}$) and effective emittance ($\epsilon$ = 0.2) in equation-2, one can obtain the equilibrium temperature T = 834 K for $Sn_{0.9}Mn_{0.1}O_{2}$ film/glass substrate. This calculated equilibrium temperature is higher than the crystallization temperature of $SnO_{2}$ [73, 78, 79] and expected to be develop within the grains of material during irradiation and facilitate (i) increase of grain size (judged by AFM), (ii) removal of micro-strain (evidenced by XRD measurements), and (iii) migration of point defects (explained by XRD and electrical measurements).\\\\
\begin{figure}
  \centering
  \includegraphics[height=7.5cm, width=9cm]{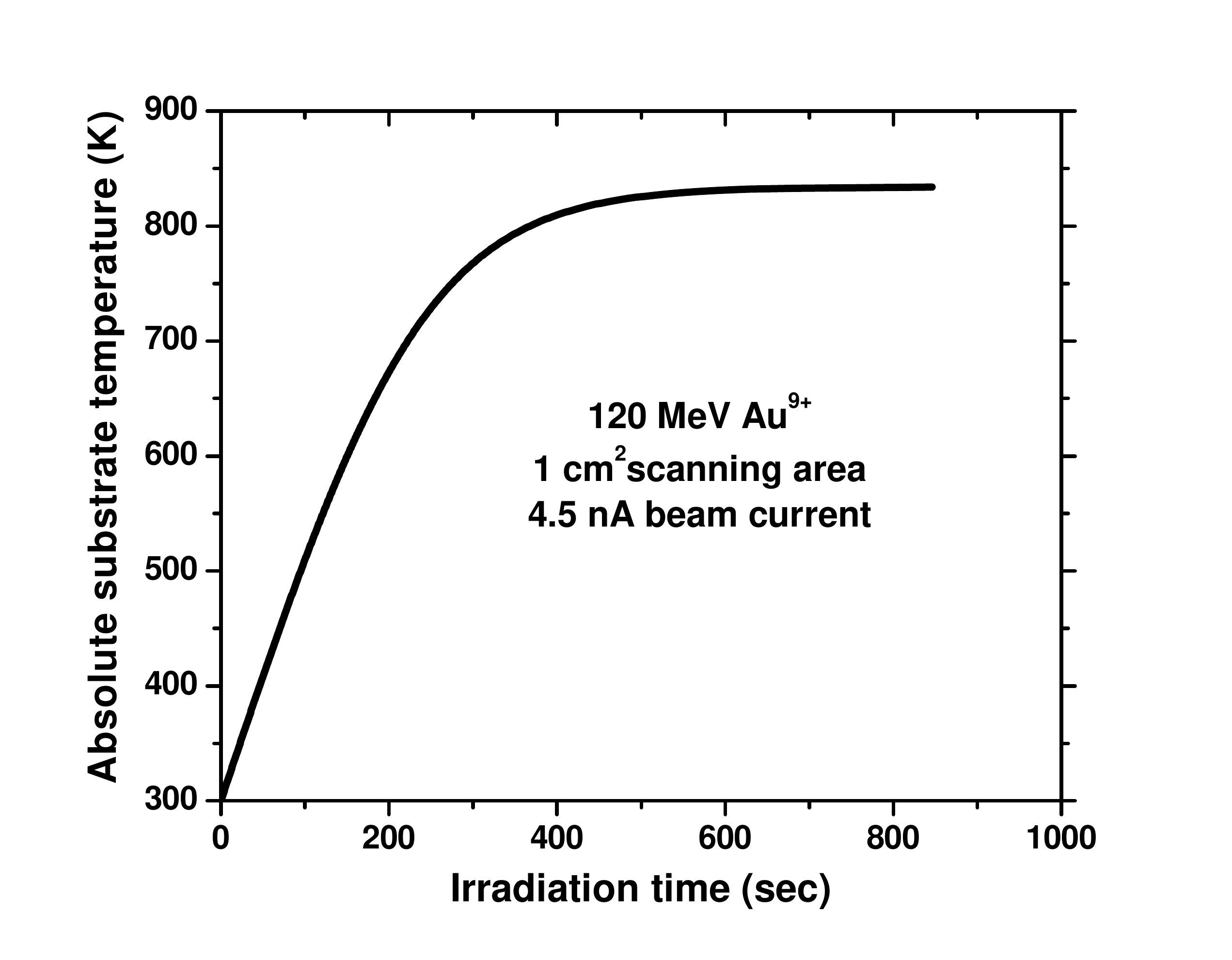}
  \caption{The transient substrate/film temperature calculated by using an input power density of 0.54 $\frac{W}{cm^{2}}$.}\label{2}
\end{figure}
The increase rate of the glass substrate temperature during irradiation can be deduced by using the following equation:
\begin{equation}\label{3}
\frac{dT}{dt} = \frac{[P - \epsilon\sigma (T^{4} - T_{s}^{4})]}{\rho wC},
\end{equation}
where C is the heat capacity, $\rho$ is the glass substrate density, and w is the substrate thickness.\\\\
Eq. (3) can be solved by separation of variable technique as follows:
\begin{equation}\label{4}
dt = \frac{dT}{a(T^{4} - b)},
\end{equation}
where a = $\frac{-\epsilon \sigma}{\rho w C}$ = -4.6304$\times$$10^{-12}$ $K^{-3}s^{-1}$ and b = $\frac{P}{\epsilon\sigma}$ + $T_{s}^{4}$ = 4.8429$\times$$10^{11}$ $K^{4}$.\\\\
The differential Eq. (4) can be numerically solved (by Wolfram language [80]) as follows:\\\\
t = $c_{1}$ - 93.002 ln (834.212 - T) + 93.002 ln (T + 834.212) + 186.004 $tan^{-1}$ (1.199 $\times$ $10^{-3}$T)\\\\
Apply initial condition: T~=~300 K at t~=~0.\\\\
t = -93.002~ln~(834.212~-~T)~+~93.002~ln~(T~+~834.212)
\begin{equation}\label{5}
~+~186.004~\tan^{-1}~(1.199~\times~10^{-3}~T)~-~134.234
\end{equation}
In this calculation, the glass substrate thickness (w), effective emittance ($\epsilon$), target material density ($\rho$), target heat capacity (C), and surrounding target temperature ($T_{s}$) are assumed to be 0.11 cm, 0.2, 2.53 $g/cm^{3}$, 0.88 J/gK, and 300 K, respectively. The irradiation fluences in the present experiment were in the range of $1 \times 10^{11}$ - $3 \times 10^{13}$ $ions/cm^{2}$, which means that the irradiation periods were in the range of 32-9600 sec when 1 $cm^{2}$ sample was used (see Table II). The results of Eq. (5) are shown in Fig. 2.  It is clear from Fig. 2 that the glass substrate temperature reaches its equilibrium value ($\sim$ 834 K) in several hundred seconds.  In the case of high fluence ($\geq$ $3 \times 10^{12}$ $ions/cm^{2}$), irradiation period is much longer than the amount of time needed to achieve equilibrium temperature. Therefore, in this case, the samples spend their almost full irradiation period at equilibrium temperature. This equilibrium temperature can influence the physical properties of samples.
\subsection{Brief review on the properties of point defects in $SnO_{2}$}
 \begin{table*}[htbp]
\renewcommand{\arraystretch}{1.4}
\centering
\caption{Properties of native or intrinsic defects in n-type $SnO_{2}$ (oxygen poor condition) [41, 42, 44-50, 84-118].}
\vspace{3mm}
\begin{tabular}{||c|c|c|c|c|c|c||}
\hline
\hline
Defect & Character & Formation &Migration& Formation  &Formation & Effect on \\
                 &  & energy& energy/temperature &probability in & probability in & lattice\\
                 &  & (Transition level)& & pristine samples& irradiated samples& \\
\hline
\hline
$V_{O}^{2+}$& Donor& Stable in  &Very high & -& - &Expansion \\
            &       & p-type  & & &  & \\
\hline
$V_{O}^{+}$& Donor& Unstable for & - & - & - & - \\
            &       & all $E_{F}$  & &  &  & \\
\hline
$V_{O}^{0}$& Neutral& Modest & Very high& Modest & Very high &Contraction \\
            &       & (Deep $\sim$ 1.8 eV)  & ($\approx 900 K$) & &  & \\
\hline
$Sn_{i}^{4+}$& Donor& Very high $\sim$ 12 eV & Low& Very low & Very low & Expansion\\
        &       & (Shallow)  &$\sim$ 0.43 eV & &  & \\
\hline
$Sn_{i}^{3+}$& Donor& Unstable & - & - & - &- \\
\hline
$Sn_{i}^{2+}$& Donor& Unstable & - & - & - &-\\
\hline
$Sn_{O}^{4+}$& Donor&  Very high & High & Very low & High & Expansion\\
             &       & (Shallow)  &($\approx 500 K$) & &  & \\
\hline
$O_{i}^{2-}$& Acceptor& High & Modest & Low & High & Expansion\\
             &       & (Deep)  &($< 500 K$) & &  & \\
\hline
$V_{Sn}^{4-}$& Acceptor& Modest in n-type & High & Modest & Very high & Expansion\\
 &       & (Deep)  &($\approx 500 K$)& (Very low @ $T > 500 K$) &  & \\
\hline
$O_{Sn}^{2-}$& Acceptor& Very high & High & Very low & High & Contraction\\
 &       & (Deep)  &($\approx 500 K$) & &  & \\
 \hline
$H_{i}^{+}$& Donor& Low & Low & Very low at& - & -\\
    &       & (Shallow)  &$\sim$ 0.57 eV &room temperature &  & \\
\hline
$H_{O}^{+}$& Donor& Low & Very high & Very high & - & Expansion\\
    &       & (Shallow) &$\sim$ 2.2 eV &  &  & \\
\hline
\hline
\end{tabular}
\end{table*}
Point defects are usually electrically active and introduce levels in the energy band gap of semiconductor. These levels involve transitions between different charge states of the same defect. Transition levels can be derived directly from the calculated formation energies [81-83]. In semiconductor there may exist two types of transition levels: shallow and deep. This distinction is on the basis of their position in the energy band with respect to valence or conduction band. For a defect to contribute in n-type conductivity, it must be stable in a positive charge state and the transition level from the positive to neutral charge state $\epsilon(+/0)$ should occur close to or above the conduction band minimum (CBM). Shallow donors are defects in which the transition level from a positive to the neutral charge state $\epsilon(+/0)$ is near or above the CBM. Theoretical studies based on density functional theory have contributed to deeper understanding of the role of point defects on the unintentional n-type conductivity in $SnO_{2}$ [41-45, 84-92]. For a long time it has been envisaged that the unintentional n-type conductivity in $SnO_{2}$ is caused by the presence of oxygen vacancies ($V_{O}$) and tin interstitials ($Sn_{i}$) [44]. However, recent first principle calculations have demonstrated that this attribution to native point defects cannot be correct [41, 42]. It has been shown that oxygen vacancies are very deep rather than shallow donors and therefore cannot contribute to n-type conductivity. According to Singh et. al. [42], the $\epsilon_{V_{O}}(2+/0)$ transitional level is located at 1.24 eV above the VBM in GGA and at 1.39 eV in GGA + U; the extrapolated value is 1.80 eV. This result distinctly shows that $V_{O}^{0}$ is a deep donor with ionization energy of 1.80 eV. For $V_{O}^{0}$, the three nearest-neighbors Sn atoms are displaced inward by 2.5\%, whereas for $V_{O}^{+}$ and $V_{O}^{2+}$, the relaxations are outward by 5.6\% and 10\% of the equilibrium Sn-O bond length, respectively.
In addition, it was found that tin interstitials ($Sn_{i}^{4+}$) and tin antisites ($Sn_{O}^{4+}$) are also unlikely causes of the unintentional n-type conductivity in as-deposited $SnO_{2}$ films [42]. The tin interstitial is a shallow donor, but it is not thermally stable. It has high formation energy about 12 eV in n-type $SnO_{2}$  and migrates through the interstitial channels along the [001] direction with an energy barrier of only 0.13 eV. This very low energy barrier implies that the Sn interstitials are highly mobile even below room temperature. Tin antisites are also shallow donors but their high formation energies make them unlikely to exist in measurable concentrations under equilibrium conditions. The dissociation barrier of $Sn_{O}^{4+}$ into $Sn_{i}^{4+}$ and $V_{O}^{0}$ is very high in n-type $SnO_{2}$. Therefore, $Sn_{O}^{4+}$ is expected to be stable at temperature of up to 500 K. This suggest that $Sn_{O}^{4+}$ may play a role in n-type conductivity under non-equilibrium conditions such as ion beam irradiation. On the other hand Sn vacancies ($V_{Sn}^{4-}$) are deep acceptors and have lowest formation energies among the native point defects under n-type conditions [42, 84-86]. They can therefore occur as a compensating defects in n-type $SnO_{2}$. In contrast, oxygen antisites ($O_{Sn}^{2-}$) have the highest formation energies among the acceptor type native defects [84-86]. Hence, it is very unlikely that oxygen antisites would be present in equilibrium. However, $O_{Sn}^{2-}$ could potentially be created under non-equilibrium conditions in n-type $SnO_{2}$. Oxygen antisites are also deep acceptors and show large off site displacements, in which the oxygen atom forms a chemical bond with one of the oxygen nearest neighbours. Oxygen interstitials $(O_{i})$ can exist either as electrically inactive split interstitials in p-type samples or as a deep acceptors at the octahedral sites in n-type samples [84-86]. In both forms oxygen interstitials have high formation energies and are therefore unlikely to be present in measurable concentrations under equilibrium conditions. A key conclusion from the recent DFT studies [41, 42, 84-88] is that native point defects cannot explain the unintentional n-type conductivity. One has to consider the role of impurities that are most likely to be present in different growth and processing environments and act as shallow donors. Hydrogen is indeed a especially ambidextrous impurity in this respect, since it is extremely difficult to detect experimentally [84, 86]. In both forms, substitutional and interstitial, hydrogen has been predicted to act as a shallow donor in n-type $SnO_{2}$ [42]. But by means of density functional calculations, it is found that interstitial hydrogen ($H_{i}^{+}$) is highly mobile and can easily diffuse out of the $SnO_{2}$ samples. The migration barrier of interstitial hydrogen is only 0.57 eV. This low barrier value makes it difficult to explain the stability of the n-type conductivity at relatively high temperature. On the other hand substitutional hydrogen ($H_{O}^{+}$) species is much more thermally stable than interstitial hydrogen and can explain stability of n-type conductivity at high temperature and its variation with partial pressure of oxygen in annealing environments [93, 94]. Both the substitutional and interstitial forms of hydrogen have low formation energies in $SnO_{2}$, indicating that they can occur in significant concentrations. Hydrogen is by no means the only possible shallow donor impurity in tin oxide, but it is a very likely candidate for an impurity that can be unintentionally incorporated and can explain observed unintentional n-type conductivity [84].
Several groups have reported on the incorporation of hydrogen in tin oxide and many have claimed that hydrogen substitutes for oxygen [41, 42, 45-50, 84-86, 95-106]. Table III lists the details of all native point defects and hydrogen impurities, based on several reported results [41, 42, 44-50, 84-118]. These summarized results will help us later to interpret experimental data in this research paper.
\subsection{X-ray diffraction studies}
\begin{figure}
  \includegraphics[height=10.4cm, width=8.3cm]{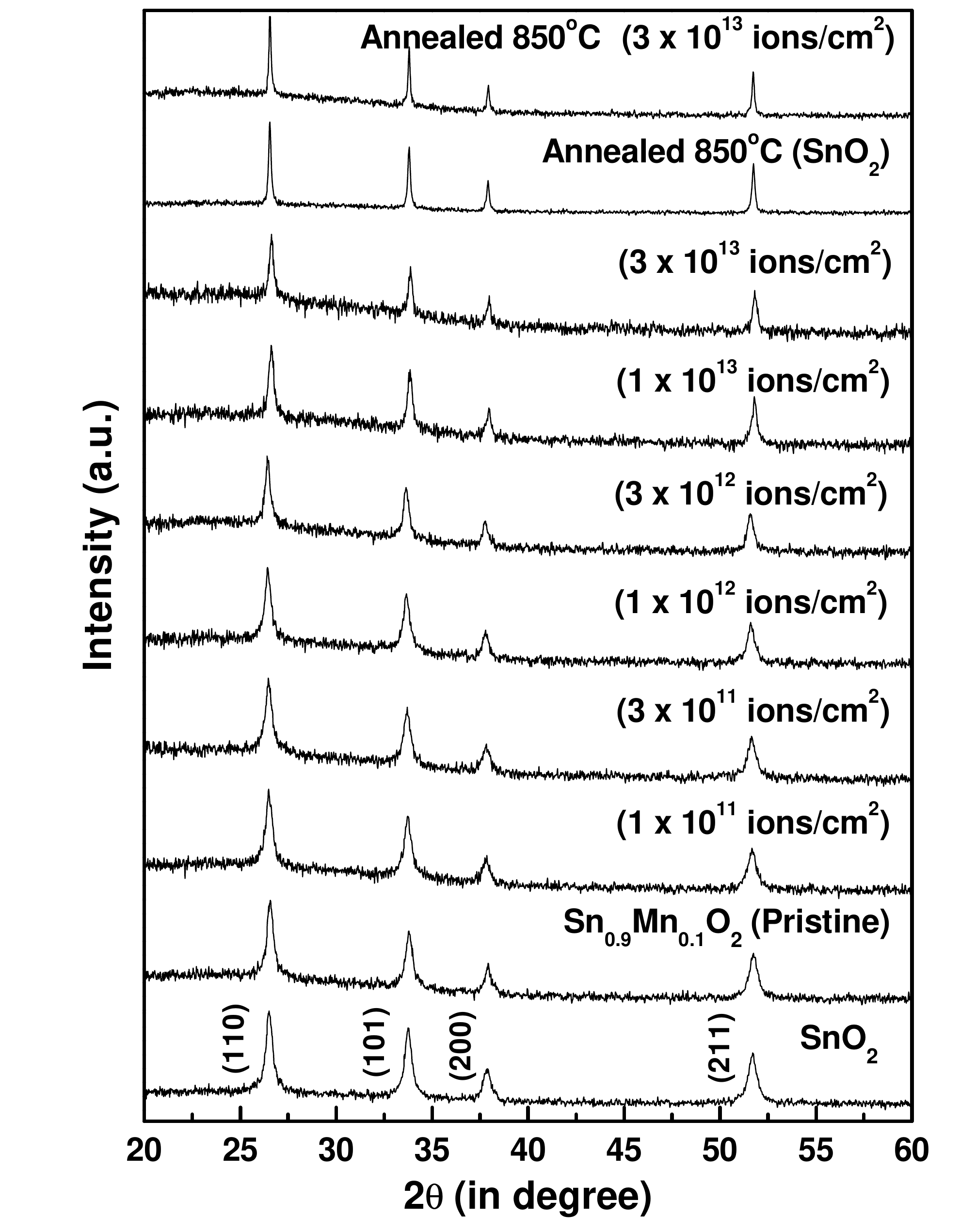}\\
  \caption{X-ray diffraction patterns of unirradiated, irradiated and annealed thin films.}\label{3}
\end{figure}
\begin{figure}
  \centering
  \includegraphics[height=6.9cm, width=8.8cm]{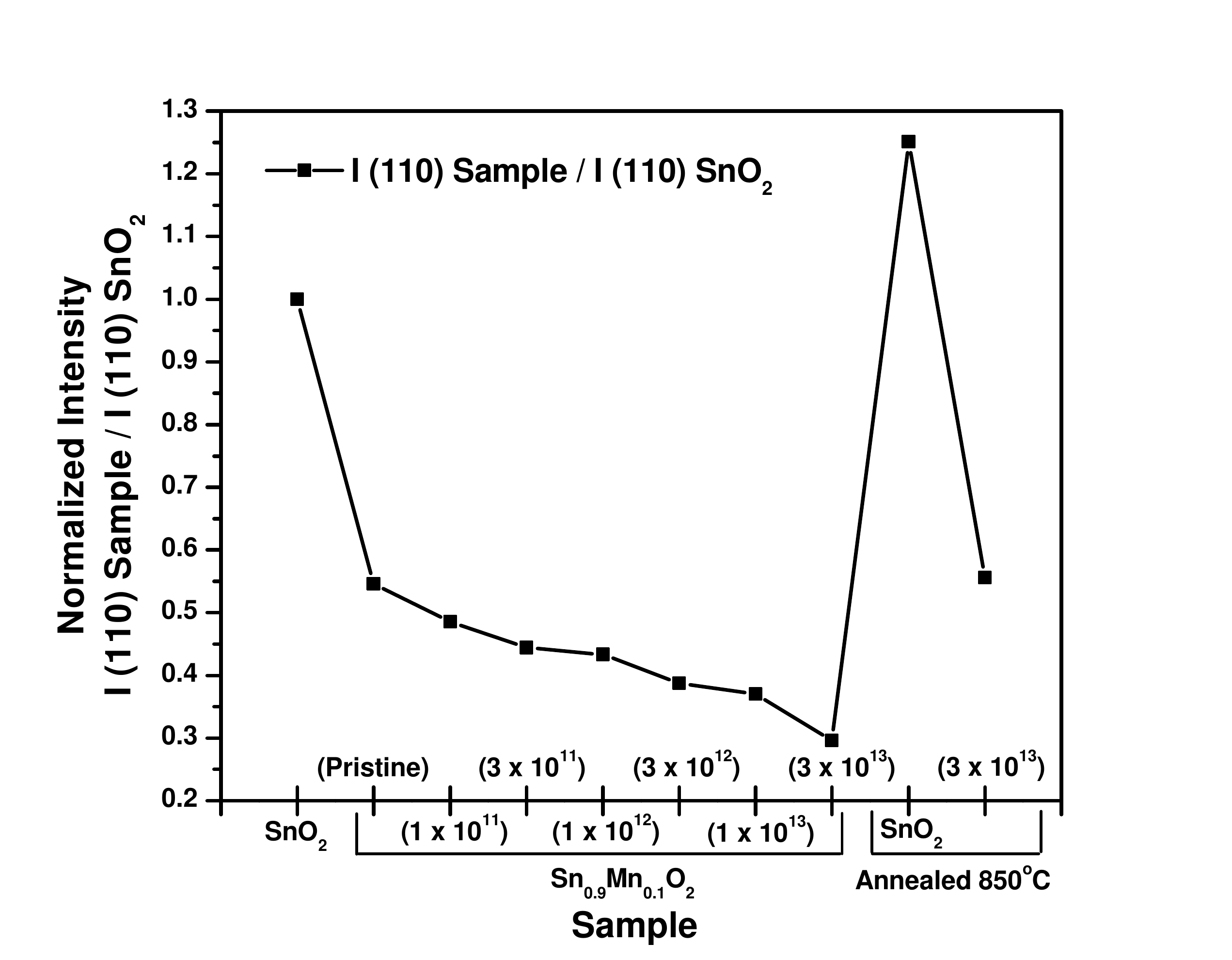}\\
  \caption{Changes in normalized intensity of respective samples.}\label{4}
\end{figure}
\begin{figure}
  \centering
  \includegraphics[height=6.9cm, width=8.8cm]{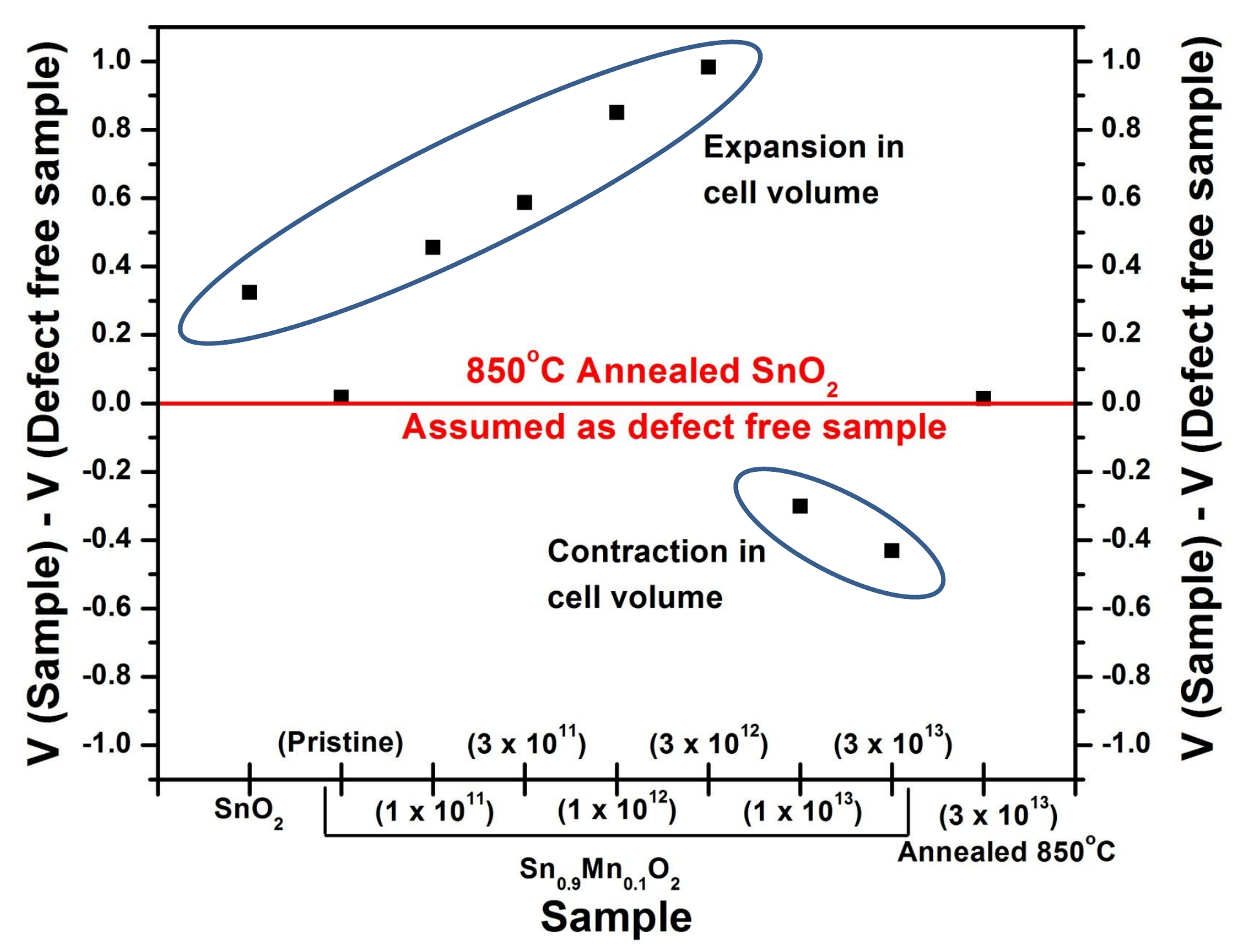}\\
  \caption{Deviations in cell volume of all samples with respect to defect free sample.}\label{5}
\end{figure}
Fig. 3 displays the x-ray diffraction patterns of unirradiated ($SnO_{2}$ and $Sn_{0.9}Mn_{0.1}O_{2}$), irradiated ($1 \times 10^{11}$ - $3 \times 10^{13}$ $ions/cm^{2}$) and annealed ($SnO_{2}$ and $3 \times 10^{13}$ $ions/cm^{2}$) samples. From the analysis of x-ray diffraction patterns, it is evident that only the peaks corresponding to the rutile-type $SnO_{2}$ phase (space group $P4_{2}/mnm$) are detected in all samples. No additional reflection peaks related to impurities, such as unreacted manganese metal, oxides or any other tin manganese phases are detected. The lack of any impurity phases shows that the Mn is evenly distributed throughout the $SnO_{2}$. The lattice parameters (a and c) and cell volume (V) are calculated by using the (110), (101) and (200) reflection planes in the XRD patterns. The complete structural analyses of the samples are given in the Table IV.
The observed contraction in cell volume from 72.1681 ${\AA}^{3}$ to 71.8616 ${\AA}^{3}$ on Mn doping, indicating the replacement of $Sn^{4+}$ by $Mn^{3+}$. The lattice parameter a is observed to decreases from 4.7491 ${\AA}$ to 4.7413 ${\AA}$, while the lattice parameter c decreases from 3.1998 ${\AA}$ to 3.1967 ${\AA}$. These results are in agreement with the fact that the difference in the ionic radius of $Mn^{3+}$ (0.58 ${\AA}$) and $Sn^{4+}$ (0.69 ${\AA}$) is rather small, leading to very small changes in the lattice constants on doping with Mn. For Mn, there are three oxidation states namely $Mn^{4+}$ (0.53 ${\AA}$), $Mn^{3+}$ (0.58 ${\AA}$) and $Mn^{2+}$ (0.82 ${\AA}$) [119]. The ionic radius of each ion in ${\AA}$ is given in brackets. But the magnitude of cell volume change indicates that the valence of the Mn ion is most probably +3 instead of +2 or +4. This speculation also finds support from the studies performed on manganese oxides showing that in its oxides, $Mn^{3+}$ prefers to be in an octahedral coordination [120], just like $Sn^{4+}$ in the rutile-type structure of tin oxide ($SnO_{2}$). It is clear from Fig. 4 that Mn substitution effects the intensity of $SnO_{2}$ peaks. The intensity of the scattered x-ray is related to structure factor and this factor is determined by the presence of bound electrons in an atom. Since manganese has less bound electrons than tin, the substitution of tin by manganese in the tin oxide lattice should yield a structure factor, which is almost half the value for tin. If there is any change in the occupation site of manganese, i.e., substitutional ($Mn_{Sn}^{3+}$) to interstitial ($Mn_{i}$), the same may be reflected as an increase in the structure factor.
\begin{table*}[htbp]
\renewcommand{\arraystretch}{1.6}
\centering
\caption{Summary of various results obtained through analysis of XRD patterns.}
\vspace{3mm}
\begin{tabular}{||c|c|r|c|c|c||}
\hline
\hline
\centering
Samples & \multicolumn{2}{c|}{Lattice parameters}& Cell volume & Defects present  &Defects \\ \cline{2-3}
  & a = b $(\dot{A})$ & c $(\dot{A})$  & $(\dot{A}^{3})$& in the samples& annealed/migrated \\ \hline \hline
  $SnO_{2}$& 4.7491 & 3.1998 & 72.1681 & $H_{O}^{+}$ & $V_{Sn}^{4-}$ \\
\hline
$Sn_{0.9}Mn_{0.1}O_{2}$& 4.7413 & 3.1967 & 71.8616 & $H_{O}^{+}$ & $V_{Sn}^{4-}$ \\
(Pristine) & &  & &  &\\
\hline
$1 \times 10^{11}$ $ions/cm^{2}$& 4.7521& 3.2016 & 72.3000 & $H_{O}^{+}$ (Partially Present)& $H_{O}^{+}$ (Partially Annealed),\\
& &  & &$V_{Sn}^{4-} > V_{O}^{0} > O_{i}^{2-} > Sn_{O}^{4+} > O_{Sn}^{2-}$ &$Sn_{i}^{4+}$\\
\hline
$3 \times 10^{11}$ $ions/cm^{2}$& 4.7551 & 3.2034 & 72.4320 & $H_{O}^{+}$ (Partially Present) & $H_{O}^{+}$ (Partially Annealed), \\
& &  & & $V_{Sn}^{4-} > V_{O}^{0} > O_{i}^{2-} > Sn_{O}^{4+} > O_{Sn}^{2-}$ & $Sn_{i}^{4+}$\\
\hline
$1 \times 10^{12}$ $ions/cm^{2}$& 4.7611 & 3.2069 & 72.6942 & $H_{O}^{+}$ (Partially Present) &$H_{O}^{+}$ (Partially Annealed), \\
& &  & & $V_{Sn}^{4-} > V_{O}^{0} > O_{i}^{2-} > Sn_{O}^{4+} > O_{Sn}^{2-}$ & $Sn_{i}^{4+}$\\
\hline
$3 \times 10^{12}$ $ions/cm^{2}$& 4.7641 & 3.2087 & 72.8267 & $H_{O}^{+}$ (Partially Present) &$H_{O}^{+}$ (Partially Annealed), \\
& &  & & $V_{Sn}^{4-} > V_{O}^{0} > O_{i}^{2-} > Sn_{O}^{4+} > O_{Sn}^{2-}$ & $Sn_{i}^{4+}$\\
\hline
$1 \times 10^{13}$ $ions/cm^{2}$& 4.7354 & 3.1905 & 71.5438 & $V_{O}^{0}$ & $H_{O}^{+}, Sn_{i}^{4+}, O_{i}^{2-},$\\
&  &  &  &  & $Sn_{O}^{4+}, O_{Sn}^{2-}, V_{Sn}^{4-}$ \\
\hline
$3 \times 10^{13}$ $ions/cm^{2}$& 4.7324 & 3.1887 & 71.4129 & $V_{O}^{0}$ & $H_{O}^{+}, Sn_{i}^{4+}, O_{i}^{2-},$\\
&  &  &  &  & $Sn_{O}^{4+}, O_{Sn}^{2-}, V_{Sn}^{4-}$ \\
\hline
Annealed $850^{o}C$& 4.7431 & 3.1935 & 71.8442 & - & $H_{O}^{+}, Sn_{i}^{4+}, O_{i}^{2-}, Sn_{O}^{4+},$ \\
($SnO_{2}$) & &  & &  & $O_{Sn}^{2-}, V_{Sn}^{4-}, V_{O}^{0}$ \\
\hline
Annealed $850^{o}C$& 4.7437 & 3.1933 & 71.8578 & - & $H_{O}^{+}, Sn_{i}^{4+}, O_{i}^{2-}, Sn_{O}^{4+},$ \\
($3 \times 10^{13}$) & &  &  & & $O_{Sn}^{2-}, V_{Sn}^{4-}, V_{O}^{0}$ \\
\hline \hline
\end{tabular}
\end{table*}

As far as we know, the migration temperatures of all possible native defects of $SnO_{2}$ are less than $700^{o}$C (see Table-III). Therefore, samples which have been annealed at a temperature of $850^{o}$C, they may be free of defects. In this article, we have assumed annealed-samples as defect-free samples (see Fig. 5). After analysis of XRD patterns, we found that the unit cell volume of as-deposited $SnO_{2}$ thin film is slightly more than that of defect free annealed $SnO_{2}$ thin film (see Fig. 5). This increase in cell volume may be due to presence of expansion-type point defects such as $H_{O}^{+}$, $V_{Sn}^{4-}$, $O_{i}^{2-}$, $Sn_{O}^{4+}$, $Sn_{i}^{4+}$. Among all these expansion-type point defects only $H_{O}^{+}$ and $V_{Sn}^{4-}$ may be responsible for the observed increase in cell volume, because formation energies of other expansion-type defects such as $O_{i}^{2-}$, $Sn_{O}^{4+}$, $Sn_{i}^{4+}$ are very high as compared to formation energies of $H_{O}^{+}$ and $V_{Sn}^{4-}$ [41, 42, 84-86]. It is impossible to create such defects of high formation energies through thermodynamic/equilibrium process. The direct and unambiguous method for introducing such defects in lattice is by high-energy ion beam irradiation. We know that tin vacancies are acceptor type point defects and play a lead role in conductivity killing (see Table-III). That's why we have prepared thin films at a temperature of $450^{o}$C, which is somewhere in between the migration temperature of tin vacancy and substitutional hydrogen. Thus, we can conclude that the slight expansion observed in cell volume of pure $SnO_{2}$ thin film is only due to $H_{O}^{+}$ point defects. According to Singh et. al. [42], $V_{O}^{0}$ defect is a contraction type, but when hydrogen does occupy the oxygen vacancy site ($V_{O}^{0}$) then resultant defect $H_{O}^{+}$ becomes expansion type, because after hydrogen substitution the resultant Sn-H bond length (2.14 ${\AA}$) is slightly more than Sn-O bond length (2.07 ${\AA}$). In a later section-III.F, we have reported the results of electrical measurements. From these measurements we can say carrier concentration of as prepared pure $SnO_{2}$ thin film is very high. According to Singh et. al. \& Janotti et. al. [42, 84-86], $H_{O}^{+}$ is also stables enough to explain this observed high unintentional n-type conductivity.

The swift heavy ion irradiation is a very effective technique to create point defects in the target material. Both expansion ($V_{Sn}^{4-}$, $O_{i}^{2-}$, $Sn_{O}^{4+}$) and contraction ($V_{O}^{0}$, $O_{Sn}^{2-}$) type of native defects can be created with the help of ion beam. These defects are created in accordance with its respective formation energy; those having less formation energy are created in large number and vice-versa. SHI irradiation can also anneal-out the defects via two ways: self-annealing and beam heating. Self-annealing is a process through which pre-existing defects of target are annealed-out. Self-annealing of pre-existing defects depends upon the ion fluence of irradiation. Almost all pre-existing defects may anneal-out at high fluence irradiation. In this process the energy transfer through ion-electron interaction makes the target abnormally excited. On the other hand, the ion beam heating process is based on the heat produced by the transfer of beam power. The power of the ion beam warms up the target to a certain equilibrium temperature in specific period of time (see Section - III.B). In the case of high fluence, irradiation period is much longer than the amount of time needed to achieve equilibrium temperature. Therefore, in this case those defects cannot be generated permanently by irradiation whose migration temperature is less than equilibrium temperature. The cell volume estimated for pristine, $1 \times 10^{11}$, $3 \times 10^{11}$, $1 \times 10^{12}$, $3 \times 10^{12}$, $1 \times 10^{13}$, $3 \times 10^{13}$ $ions/cm^{2}$ irradiated films are 71.8616, 72.3000, 72.4320, 72.6942, 72.8267, 71.5438 and 71.4129 ${\AA}^{3}$, respectively. Due to self-annealing of pre-existing expansion-type $H_{O}^{+}$ defects, the irradiated samples should show a reduction in the unit cell volume. But here, instead of decrease in cell volume, an increase has been observed upto $3 \times 10^{12}$ $ions/cm^{2}$ (see Fig. 5). In spite of self-annealing, ion beam can also generate defects in accordance with its formation energy. As we know, the expansion-type $V_{Sn}^{4-}$ defects have lowest formation energy among the all native point defects ($V_{Sn}^{4-}$, $V_{O}^{0}$, $O_{i}^{2-}$, $Sn_{O}^{4+}$, $O_{Sn}^{2-}$) under n-type conditions (see Table III). Therefore, after irradiation, they may be present in large numbers as compared to other native defects and can explain the observed increase in cell volume. After the irradiation fluence of $3 \times 10^{12}$ $ions/cm^{2}$, we have observed a decrease in the cell volumes of samples (see Fig. 5). This decrease in the cell volumes can be explained in terms of ion beam heating phenomena. In case of high fluence (such as $1 \times 10^{13}$ or $3 \times 10^{13}$ $ions/cm^{2}$), material spends almost entire time during irradiation at equilibrium temperature ($\sim$ 834 K). So possibly in such case, those point defects cannot be created permanently through irradiation whose migration temperature is less than equilibrium temperature. Among the all possible defects ($V_{Sn}^{4-}$, $V_{O}^{0}$, $O_{i}^{2-}$, $Sn_{O}^{4+}$, $O_{Sn}^{2-}$) only contraction-type $V_{O}^{0}$ defect has greater migration temperature ($\sim$ 900 K) than equilibrium substrate/film temperature ($\sim$ 834 K) [42, 84-86]. Therefore, only contraction-type $V_{O}^{0}$ defects may be present in high fluence irradiated samples and can explain the observed contraction in cell volume. We have also analyzed the effect of annealing on the cell volumes of irradiated and pure $SnO_{2}$ samples. It should be noted that the cell volumes of both annealed films (irradiated at $3 \times 10^{13}$ $ions/cm^{2}$ and pure $SnO_{2}$) are almost equal. This observation confirms the creation of point defects in irradiated materials. If we keep annealing temperature of irradiated samples higher than the migration temperature of all possible defects (approximately $700^{o}$) then all defects present in the irradiated samples becomes migrate and consequently samples acquire defect free state.
\subsection{Texture coefficient, crystallite size, micro and macro-strain}
\begin{table*}[htbp]
\centering
\renewcommand{\arraystretch}{1.9}
\caption{The Texture coefficient C(hkl), the degree of preferential orientation $\sigma$, crystallite size, non-uniform and uniform strain for all the samples.}
\vspace{3mm}
\begin{tabular}{||c|c|r|c|c|c|c|c|c|c||}
\hline
\hline
\centering
 & \multicolumn{4}{c|}{Texture coefficient C(hkl)}& & Crystallite  & Non-Uniform & \multicolumn{2}{c||}{Uniform Strain}\\ \cline{2-5}\cline{9-10}
 Samples & (110) & (101) & (200) & (211) & $\sigma$ & Size (nm) & Strain& a - direction & c - direction \\ \hline \hline
  $SnO_{2}$& 0.792& 0.865& 1.442& 0.901& 0.258 & 27 & $12.5\times10^{-4}$& $1.265\times10^{-3}$ & $1.973\times10^{-3}$\\
\hline
$Sn_{0.9}Mn_{0.1}O_{2}$& 0.814& 0.783& 1.504& 0.899& 0.294 & 29 & $12.7\times10^{-4}$ & $-0.379\times10^{-3}$ & $1.002\times10^{-3}$\\
(Pristine)& & & & & & & & & \\
\hline
$1\times 10^{11}$ $ions/cm^{2}$& 0.844& 0.826& 1.498& 0.833& 0.287 & 26 & $11.4\times10^{-4}$ & $1.897\times10^{-3}$ & $2.536\times10^{-3}$\\
\hline
$3\times 10^{11}$ $ions/cm^{2}$& 0.827& 0.824& 1.476& 0.873& 0.276 & 26 & $10.8\times10^{-4}$ & $2.529\times10^{-3}$ & $3.100\times10^{-3}$\\
\hline
$1\times 10^{12}$ $ions/cm^{2}$& 0.820& 0.812& 1.515& 0.853& 0.298 & 29 & $8.71\times10^{-4}$ & $3.795\times10^{-3}$ & $4.196\times10^{-3}$\\
\hline
$3\times 10^{12}$ $ions/cm^{2}$& 0.850& 0.816& 1.480& 0.855& 0.277 & 34 & $6.24\times10^{-4}$ & $4.427\times10^{-3}$ & $4.759\times10^{-3}$\\
\hline
$1\times 10^{13}$ $ions/cm^{2}$& 0.743& 0.813& 1.473& 0.972& 0.285 & 39 & $3.77\times10^{-4}$ & $-1.623\times10^{-3}$ & $-0.939\times10^{-3}$\\
\hline
$3\times 10^{13}$ $ions/cm^{2}$& 0.739& 0.719& 1.622& 0.920& 0.367 & 45 & $1.15\times10^{-4}$ & $-2.256\times10^{-3}$ & $-1.503\times10^{-3}$\\
\hline
Annealed $850^{o}C$& 0.828& 0.795& 1.460& 0.916& 0.269 & 121 & $2.51\times10^{-4}$ & - & -\\
($SnO_{2}$)& & & & & & & & & \\
\hline
Annealed $850^{o}C$& 0.849& 0.839& 1.424& 0.887& 0.246 & 139 & $0.49\times10^{-4}$ & $0.126\times10^{-3}$ & $-0.062\times10^{-3}$\\
($3 \times 10^{13}$)& & & & & & & & & \\
\hline
\hline
\end{tabular}
\end{table*}
The preferential orientation of the crystallites in the films was studied by calculating the texture coefficient C(hkl) of each XRD peak using the equation [121, 122]:
\begin{equation}\label{6}
  C(hkl) = \frac{N(I(hkl)/I_{o}(hkl))}{\sum (I(hkl)/I_{o}(hkl))}
\end{equation}
where $C (hkl)$ is the texture coefficient of the plane (hkl), I(hkl) is the measured integral intensity, $I_{o}(hkl)$ is the JCPDS standard integral intensity for the corresponding powder diffraction peak (hkl), and N is the number of reflections observed in the x-ray diffraction pattern. C(hkl) is unity for each XRD peak in the case of a randomly oriented film and values of C(hkl) greater than unity indicate preferred orientation of the crystallites in that particular direction. The degree of preferred orientation $\sigma$ of the film as a whole can be evaluated by estimating the standard deviation of all the calculated C(hkl) values [123]:
\begin{equation}\label{7}
  \sigma = \sqrt{\frac{\sum [C(hkl) - C_{o}(hkl)]^{2}}{N}}
\end{equation}
where $C_{o}(hkl)$ is the texture coefficient of the powder sample which is always unity.
The zero value of $\sigma$ indicates that the crystallites in the film are oriented randomly. The higher value of $\sigma$ indicates that the crystallites in the film are oriented preferentially [123]. The texture coefficient C(hkl) of all the XRD peaks along with the value of $\sigma$ for each film is given in Table V. It can be seen that the plane (200) has a high texture coefficient for all the films. The thin film attained maximum preferential orientation ($\sigma$ = 0.367) when the fluence value is $3 \times 10^{13}$ $ions/cm^{2}$. However, it should be highlighted that none of the films possess a significant preferential orientation since the value of $\sigma$ is less than unity for all the films.

In polycrystalline materials, two types of lattice strain can be encountered: uniform strain (macro-strain) and non-uniform strain (micro-strain). Uniform strain causes the unit cell to contract/expand in an isotropic way. This simply leads to a change in the lattice parameters and shift of the x-ray diffraction peaks. There is no peak broadening associated with this type of lattice strain. Moreover, non-uniform strain leads to systematic shifts of atoms from their ideal positions and to peak broadening [124, 125]. In the case of nano-materials where crystallite size also plays a role in peak broadening the two effects strain and size broadening overlap each other. In such cases the micro-strain or crystallite size can only be determined by separating the two effects. However, the peak width derived from micro-strain varies as $\tan\theta$, whereas crystallite size varies as $1/\cos\theta$. This difference in behaviour as a function of $2\theta$ enables one to discriminate between the strain and size effects on peak broadening. Williamson-Hall (W-H) analysis is a simplified integral breadth method where micro-strain-induced and crystallite size-induced broadening are de-convoluted by considering the peak width as a function of $2\theta$ [126-129]. Williamson and Hall assumed that both strain and size broadened profiles are Lorentzian [127]. Based on this assumption a mathematical relation was established between the integral breadth $(\beta)$, the lattice micro-strain $(\varepsilon)$ and volume weighted average crystallite size (D) as follows:
\begin{figure}
  \centering
  \includegraphics[height=8.55cm, width=9cm]{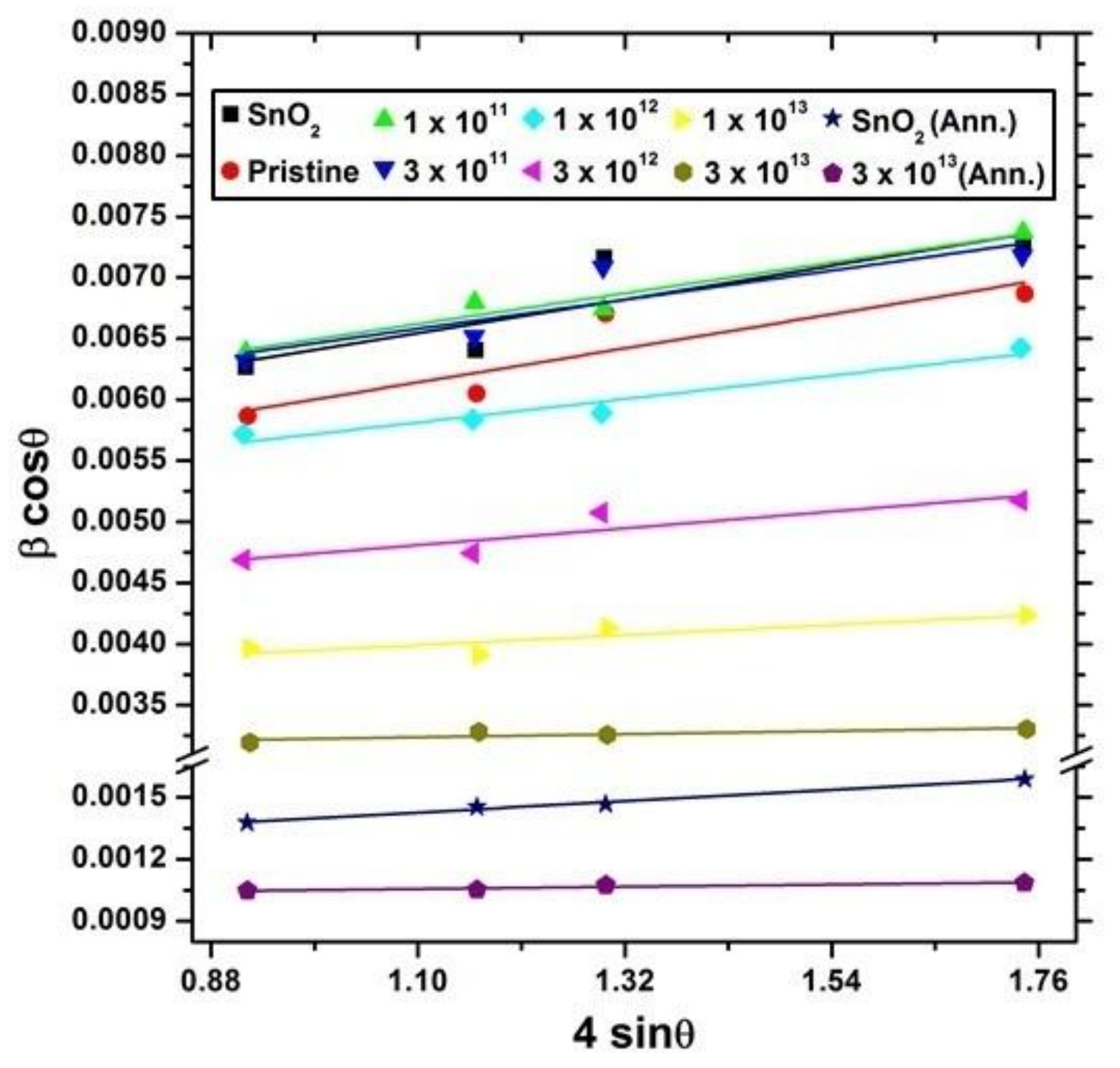}\\
  \caption{Williamson-Hall plots of all samples.}\label{6}
\end{figure}
\begin{equation}\label{8}
  \beta \cos\theta = \frac{k\lambda}{D} + 4\varepsilon \sin\theta
\end{equation}
Where k is a constant equal to 0.94, $\lambda$ is the wavelength of the radiation (1.5405 ${{\AA}}$ for $CuK_{\alpha}$ radiation), and $\theta$ is the peak position. The instrumental resolution in the scattering angle $2\theta$, $\beta_{inst}$, was determined by means of a standard crystalline silicon sample and approximated by
\begin{center}
$\beta_{inst}$ = $9\times10^{-6}$ $(2\theta)^{2}$ - 0.0005 $(2\theta)$ + 0.0623
\end{center}
Finally, the integral breadth $\beta$ without instrumental contribution was obtained according to the relation:
\begin{equation}\label{9}
  \beta  = \beta_{measured} - \beta_{instrumental}
\end{equation}

Eq. 8 represents the general form of a straight line $y = mx + c$. The plot between $\beta \cos\theta$ and $4 \sin\theta$ gives a straight line having slope $\varepsilon$ and intercept $k\lambda/D$. The values of crystallite size and micro-strain can be obtained from the inverse of intercept and the slope of the straight line, respectively. The Williamson-Hall plots for all the samples are given in Fig. 6 and the results extracted from these plots are listed in Table V. Although W-H analysis is an averaging method apart from TEM imaging, it still holds a dominant position in crystallite size determination. It is clear from intercept values that SHI irradiation does not influence the crystallite size much in the low fluence regime, but causes a significant increase at the fluence of $1 \times 10^{13}$ and $3 \times 10^{13}$ $ions/cm^{2}$. The crystallite size of the target changes from 29 nm to 45 nm after irradiation at the fluence of $3 \times 10^{13}$ $ions/cm^{2}$. Such significant change in the crystallite size at higher fluence may be due to ion-beam heating effect. In ion beam heating process the power of ion beam warms up the target to a certain equilibrium temperature in specific period of time. In the case of high fluence ($>$ $1 \times 10^{12}$ $ions/cm^{2}$) material spends almost entire time during irradiation at equilibrium temperature of $\sim$ 834 K and feels annealing like atmosphere for whole irradiation period (see section-III.B).

Furthermore, we have calculated macro-strains of all the samples along a and c directions with the help of the following relations:-
\begin{equation}\label{10}
  \varepsilon^{'}_{a} = \frac{(a - a_{o})}{a_{o}}
\end{equation}
\begin{equation}\label{11}
  \varepsilon^{'}_{c} = \frac{(c - c_{o})}{c_{o}}
\end{equation}
Where $a_{o}$ = 4.7431 ${\AA}$ and $c_{o}$ = 3.1935 ${\AA}$ are the lattice parameters of the defect free sample (annealed $SnO_{2}$ thin film). The obtained values of $\varepsilon^{'}_{a}$ and  $\varepsilon^{'}_{c}$ for all the samples are listed in Table V.

It is important to note that the Williamson-Hall method probes the micro-strain induced by dislocation inside the crystalline grains, i.e., opposite to a macro-strain that causes a shift in the Bragg peak position. By inspection of the W-H plots, it appears that the value of micro-strain for pristine film is $12.7 \times 10^{-4}$ and is reduced to $11.4 \times 10^{-4}$, $10.8 \times 10^{-4}$, $8.71 \times 10^{-4}$, $6.24 \times 10^{-4}$, $3.77 \times 10^{-4}$, $1.15 \times 10^{-4}$ for films irradiated with the fluence of $1 \times 10^{11}$, $3 \times 10^{11}$, $1 \times 10^{12}$, $3 \times 10^{12}$, $1 \times 10^{13}$, and $3 \times 10^{13}$ $ions/cm^{2}$, respectively. The reduction of micro-strain after irradiation is indicative of the improvement in crystallinity due to self-annealing of pre-existing irregular lattice defects such as dislocations, faults etc. Practically, it has been found that the higher the ion fluence, more is the heat generated, and hence smaller the micro-strain. On the other hand the macro-strain of the target is increased after irradiation (see Table V). The increase in macro-strain may be due to the formation of fresh regular expansion/contraction type point defects. It can be clearly seen from the Table V, the macro-strain is of tensile type upto $3 \times 10^{12}$ $ions/cm^{2}$ fluence and after that it is of compressive type. That is because before $1 \times 10^{13}$ $ions/cm^{2}$ fluence the density of expansion type defects is more as compare to contraction type defects and it is vice versa for $1 \times 10^{13}$ and $3 \times 10^{13}$ $ions/cm^{2}$ fluence (already discussed in the XRD section). The micro-strain for the unirradiated pure $SnO_{2}$ sample is measured as 12.5 $\times$ $10^{-4}$, which changes to 2.51 $\times$ $10^{-4}$ after annealing at $850^{o}$ (see Table V). From comparing the micro-strain of pure $SnO_{2}$ samples annealed at $850^{o}$C with samples irradiated at $3 \times 10^{13}$ $ions/cm^{2}$ fluence, we find that high fluence irradiation more efficiently releases micro-strain from the samples. The macro-strain was also found to be negligibly small for the annealed samples (pure $SnO_{2}$ and irradiated at $3 \times 10^{13}$ $ions/cm^{2}$ fluence). The reduction in micro- and macro-strain values after thermal annealing may be due to the migration of lattice defects at $850^{o}$C. From W-H plots, it is clear that the grain growth has occurred on thermal annealing at $850^{o}$C.

\subsection{Electrical properties}
The electrical properties of the films were estimated by resistivity and Hall effect measurements made at room temperature. The room temperature results are presented for all measured films in Table VI. The pure $SnO_{2}$ film shows the best combination of electrical properties as follows: resistivity ($\rho$) of 2.41 $\times$ $10^{-3}$ $\Omega$ cm, carrier concentration (n) of 2.735 $\times$ $10^{20}$ $cm^{-3}$, and mobility ($\mu$) of 9.497 $cm^{2}V^{-1}s^{-1}$.
The lower resistivity in the as-deposited $SnO_{2}$ film may be due to the presence of substitutional hydrogen ($H_{O}^{+}$). First principle calculations have provided evidence that usual suspects such as oxygen vacancy $V_{O}$ and tin interstitial $Sn_{i}$ are actually not responsible for n-type conductivity in majority of the cases [41, 42]. These calculations indicate that the oxygen vacancies are a deep donor, whereas tin interstitials are too mobile to be stable at room temperature [42]. Recent first principle calculations have drawn attention on the role of donor impurities in unintentional n-type conductivity [41, 42, 84-88]. Hydrogen is indeed a especially ambidextrous impurity in this respect, since it is extremely difficult to detect experimentally [41, 42, 84]. By means of density functional calculations it has been shown that hydrogen can substitute on an oxygen site and has a low formation energy and act as a shallow donor [41, 42, 84]. Hydrogen is by no means the only possible shallow donor impurity in tin oxide, but it is a very likely candidate for an impurity that can be unintentionally incorporated and can explain observed unintentional n-type conductivity [84]. Tin vacancies have the lowest formation energy among the native point defects under n-type condition [41, 42, 84]. Therefore they may exist as compensating defects in n-type $SnO_{2}$ sample (see Section III.C). But if we make thin films at a specific temperature somewhere in between migration temperature of tin vacancy and substitutional hydrogen, we find that they are highly conducting ($n\geq10^{20}$ $cm^{-3}$). In the present investigation, we have optimised substrate temperature  for the deposition of $SnO_{2}$ thin films at 450$^{o}$ C.  This substrate temperature is much higher than the migration temperature of tin vacancy. Thus we can say that the n-type conductivity in $SnO_{2}$ samples may be caused by $H_{O}^{+}$ defects. On the other hand, the resistivity of the thermally annealed films shows an insulating behavior. In the thermally annealed film, the transformation towards stoichiometry leads to an increase in the resistivity as expected for a metal-oxide semiconductor. Thermal annealing of as-deposited $SnO_{2}$ films was carried out beyond the migration temperature of $H_{O}^{+}$ defects ($\sim$ 900K) to confirm the role of $H_{O}^{+}$ defects on the film resistivity. The electrical resistivity of 3.63 $\times$ $10^{-3}$ $\Omega$ cm obtained from the pristine $Sn_{0.9}Mn_{0.1}O_{2}$ films is increased to 49.6 $\Omega$ cm for the films irradiated with 3 $\times$ $10^{12}$ $ions/cm^{2}$ fluence (see Fig. 7). The value of carrier concentration for pristine sample is 1.927 $\times$ $10^{20}$ $cm^{-3}$ and is reduced to 9.658 $\times$ $10^{18}$, 3.283 $\times$ $10^{18}$, 7.492 $\times$ $10^{16}$, 1.831 $\times$ $10^{16}$ $cm^{-3}$ for films irradiated with the fluence of 1 $\times$ $10^{11}$, 3 $\times$ $10^{11}$, 1 $\times$ $10^{12}$, and 3 $\times$ $10^{12}$ $ions/cm^{2}$, respectively (see Fig. 8). On the other hand, the mobility of $Sn_{0.9}Mn_{0.1}O_{2}$ films is decreased from the original value of 8.929 $cm^{2}V^{-1}s^{-1}$ to 6.875 $cm^{2}V^{-1}s^{-1}$ after irradiation with fluence of 3 $\times$ $10^{12}$ $ions/cm^{2}$ (see Fig. 8). Above a typical fluence of 3 $\times$ $10^{12}$ $ions/cm^{2}$, electrical measurements show that the sheet resistance of irradiated film is of the order of 10 M$\Omega$/$\Box$ (see Fig. 7). SHI irradiation is very effective method to keep the sample at very high temperature through electron-phonon coupling in a short span of time. Due to this self annealing effect of irradiation, the concentration of pre-existing defects ($H_{O}^{+}$) decreases with increasing ion fluence. Along with this self annealing of pre-existing defects, some fresh electron donor ($Sn_{O}^{4+}$) and electron killer ($V_{Sn}^{4-}$, $O_{i}^{2-}$, $O_{Sn}^{2-}$) type defects can be formed through irradiation in the target.
\begin{figure}
  \centering
  \includegraphics[height=7.4cm, width=9.3cm]{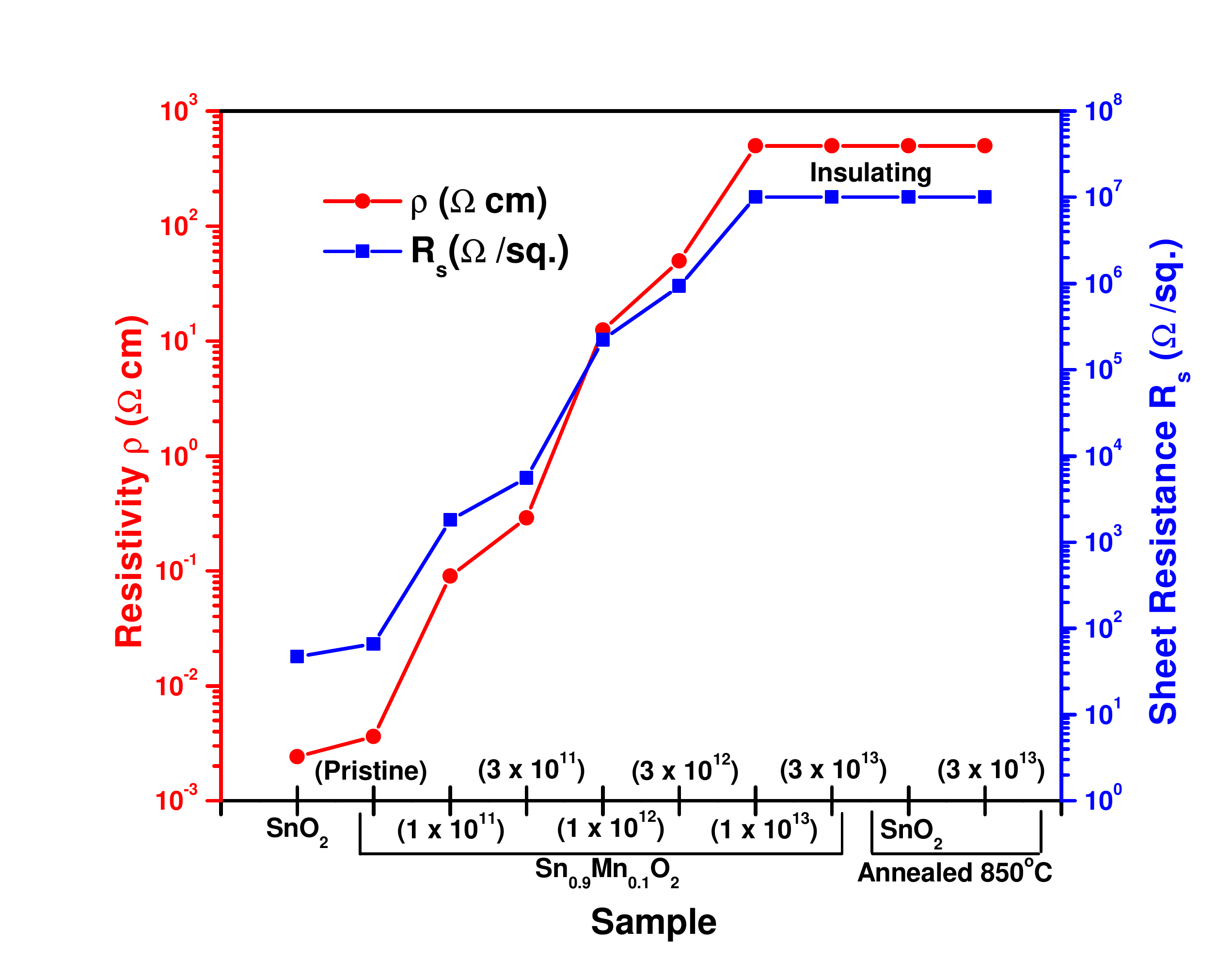}\\
  \caption{Changes in the electrical resistivity $\rho$ and sheet resistance $R_{s}$ of respective samples. Here x-axis is not scaled for better clarity.}\label{7}
\end{figure}
\begin{figure}
  \centering
  \includegraphics[height=7.4cm, width=9.3cm]{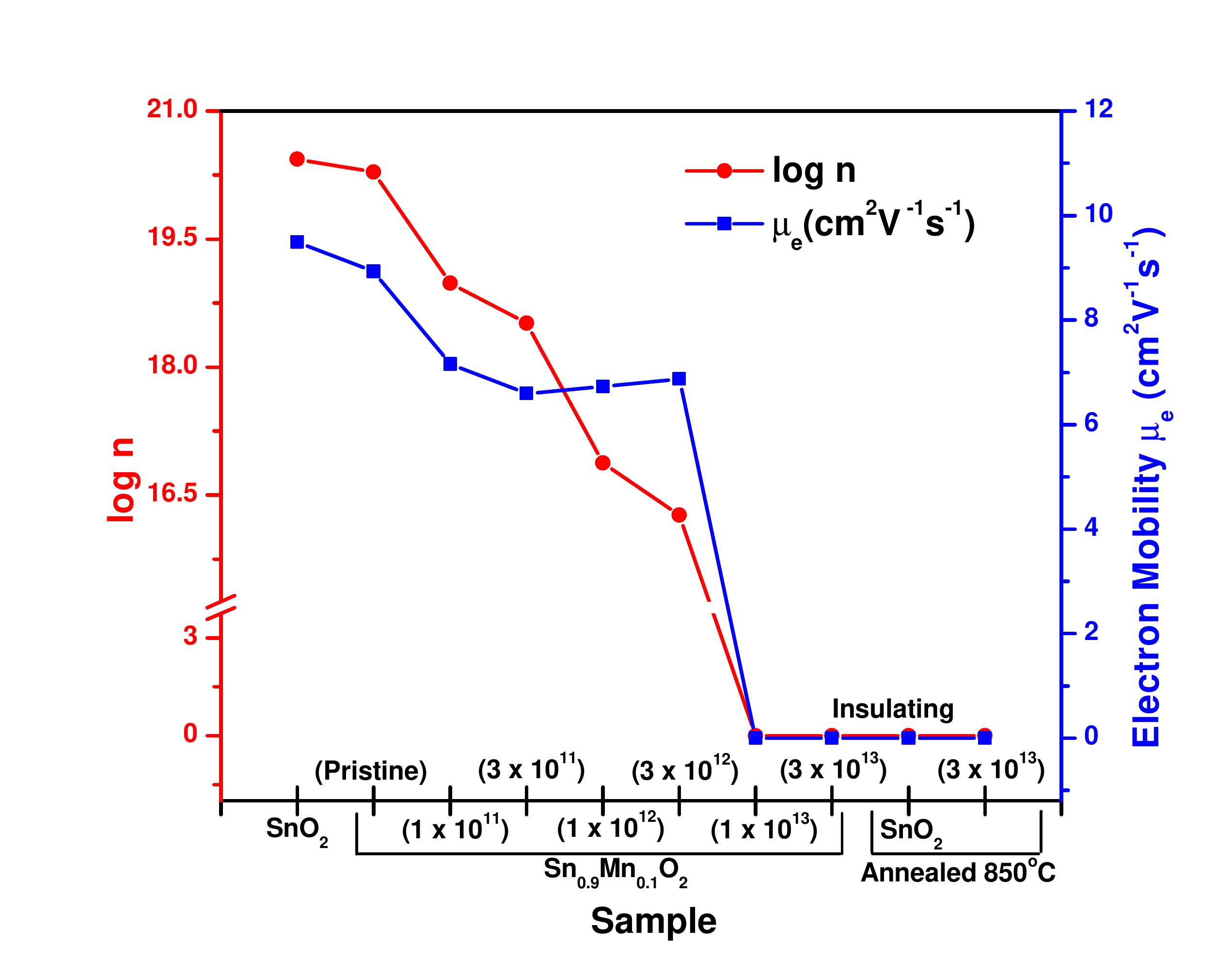}\\
  \caption{Changes in the free carrier concentration n and electron mobility $\mu_{e}$ of respective samples. Here x-axis is not scaled for better clarity.}\label{8}
\end{figure}
\begin{figure}
  \centering
  \includegraphics[height=6.95cm, width=9cm]{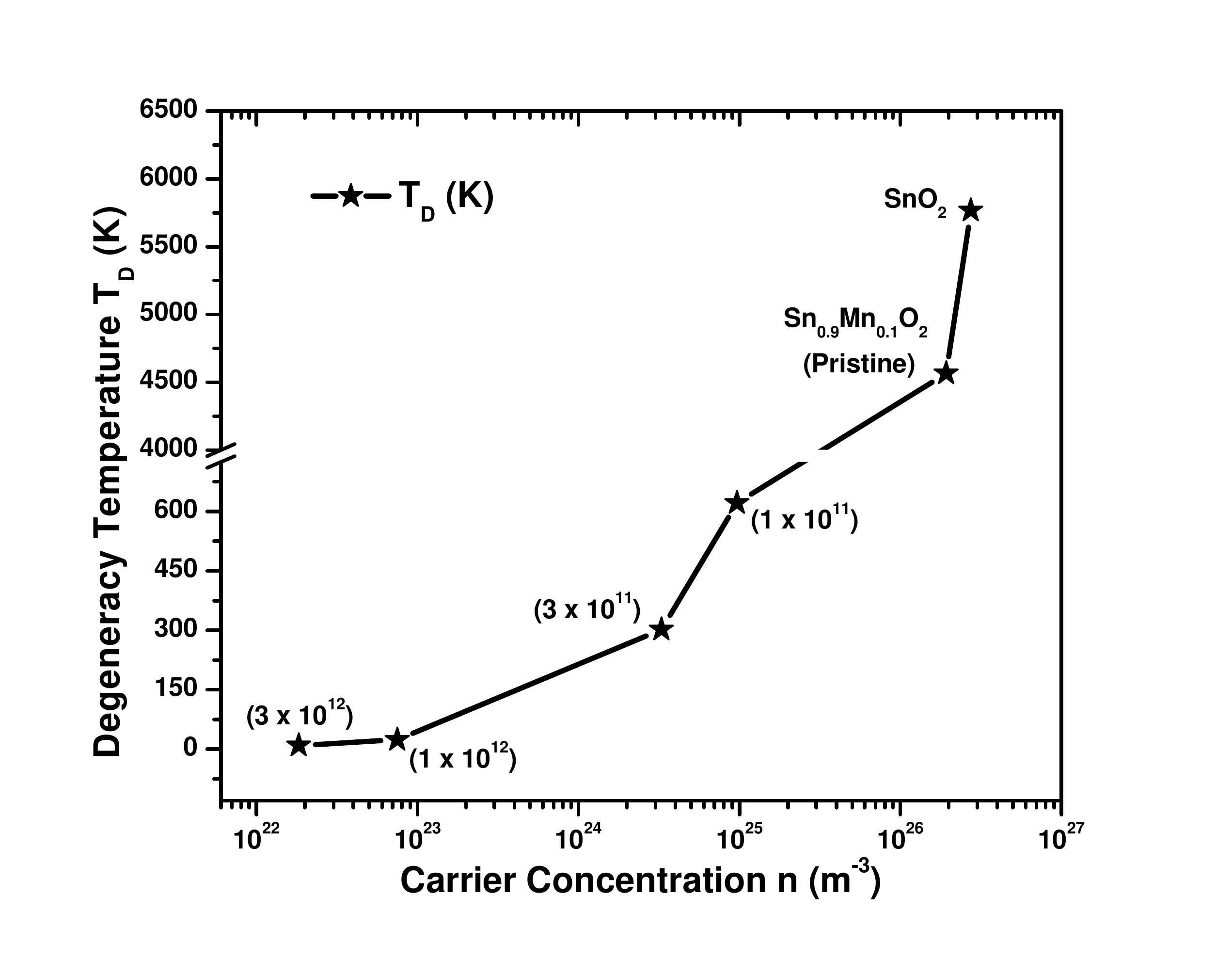}\\
  \caption{Degeneracy temperature as a function of carrier concentration.}\label{9}
\end{figure}
These fresh defects are created in accordance with its respective formation energy; those having less formation energy are created in large number and vice-versa. The variation in conductivity of irradiated samples may be caused by these mixed effects. The point defects newly created by ion beam irradiation can also be annealed out of the sample at high fluence ($\geq$ $1\times10^{13}$ $ions/cm^{2}$). In the case of high fluence, target sample spends almost entire time during irradiation at equilibrium temperature of $\sim$ 834 K (see Section - III.B) . Therefore in that case all those newly created defects may anneal out of the sample whose migration temperature is less than the equilibrium temperature. Among the all possible defects only $V_{O}^{0}$ defect has a higher migration temperature ($\sim$ 900 K) than the equilibrium temperature ($\sim$ 834 K) [42, 84-86]. Therefore other newly created defects $(V_{Sn}^{4-}, O_{i}^{2-}, Sn_{O}^{4+}, O_{Sn}^{2-})$ except $V_{O}^{0}$ may anneal out of the sample during high fluence irradiation. With the help of these arguments one can understand the insulating behaviour of high fluence irradiated samples.
\begin{table*}[htbp]
\centering
\renewcommand{\arraystretch}{1.5}
\caption{Electrical parameters for all the samples. }
\vspace{3mm}
\begin{tabular}{||c|c|c|c|c|c|c|c||}
\hline
\hline
\centering
 & Film & Sheet& & Carrier & Degeneracy& \multicolumn{2}{c||}{Carrier mobility}\\
 &thickness &resistance & Resistivity & concentration& temperature&\multicolumn{2}{c||}{$\mu$ ($cm^{2}V^{-1}s^{-1}$)}\\ \cline{7-8}
 Samples &  (nm)& $R_{s}$ & $\rho$ ($\Omega$ cm) & n ($cm^{-3}$)& $T_{D}$ (K)&Observed & Calculated\\ \hline \hline
  $SnO_{2}$& 510& 47 $\Omega/\square$& 2.41$\times10^{-3}$& $2.735\times10^{20}$& 5767 & 9.497 & 10.346\\
\hline
$Sn_{0.9}Mn_{0.1}O_{2}$& 550& 66 $\Omega/\square$& 3.63$\times10^{-3}$& $1.927\times10^{20}$& 4566 & 8.929 & -\\
(Pristine)& & & & & & & \\
\hline
$1\times 10^{11}$ $ions/cm^{2}$& 500& 1.81 $K\Omega/\square$& 9.03$\times10^{-2}$& $9.658\times10^{18}$& 621 & 7.163 & -\\
\hline
$3\times 10^{11}$ $ions/cm^{2}$& 523& 5.53 $K\Omega/\square$& 0.289& $3.283\times10^{18}$& 302 & 6.597 & -\\
\hline
$1\times 10^{12}$ $ions/cm^{2}$& 555& 0.223 $M\Omega/\square$& 12.4& $7.492\times10^{16}$& 24 & 6.724 & -\\
\hline
$3\times 10^{12}$ $ions/cm^{2}$& 530& 0.936 $M\Omega/\square$& 49.6& $1.831\times10^{16}$& 10 & 6.875 & -\\
\hline
$1\times 10^{13}$ $ions/cm^{2}$& 515& $\approx$10 $M\Omega/\square$& -& -& - & - & -\\
\hline
$3\times 10^{13}$ $ions/cm^{2}$& 505& $\approx$10 $M\Omega/\square$& -& -& - & - & -\\
\hline
Annealed $850^{o}C$& 510& $\approx$10 $M\Omega/\square$& -& -& - & - & -\\
($SnO_{2}$)& & & & & & &\\
\hline
Annealed $850^{o}C$& 505& $\approx$10 $M\Omega/\square$& -& -& - & - & -\\
($3 \times 10^{13}$ $ions/cm^{2}$)& & & & & & & \\
\hline
\hline
\end{tabular}
\end{table*}

The temperature dependence of electrical resistivity in the range 30-200$^{o}C$ indicates that the pure and Mn doped $SnO_{2}$ (pristine) films are degenerate semiconductors. The film degeneracy was further confirmed by evaluating degeneracy temperature of the electron gas $T_{D}$ by the expression [130, 131]:
\begin{equation}\label{12}
  k_{B}T_{D} \simeq (\frac{\hbar^{2}}{2m^{*}})(3\pi^{2}n)^\frac{2}{3} = E_{F},
\end{equation}
where $m^{*}$ is the reduce effective mass and n is the electron concentration. The degeneracy temperature of all investigated samples is clearly displayed in  Fig. 9. It can be seen that $T_{D}$ of $SnO_{2}$ and $Sn_{0.9}Mn_{0.1}O_{2}$ (pristine) films are well above room temperature, both at around 5000 K.
Here, We have tried to identify the main scattering mechanisms that influence the mobility of pure $SnO_{2}$ films. There are many scattering mechanisms such as grain-boundary scattering, domain scattering, surface scattering, interface scattering, phonon scattering (lattice vibration), neutral, and ionized impurity scattering which may influence the mobility of the films [132, 133]. The interaction between the scattering centres and the carriers determines the actual value of the mobility of the carriers in the samples.  In the interpretation of the mobility obtained for pure $SnO_{2}$ films, one has to deal with the problem of mixed scattering of carriers. To solve this problem, one has to identify the main scattering mechanism and then determine their contributions. The pure $SnO_{2}$ films prepared here are polycrystalline. They are composed of grains joined together by grain boundaries, which are transitional regions between different orientations of neighboring grains. These boundaries between grains play a significant role in the
scattering of charge carriers in polycrystalline thin films. The grain boundary scattering has an effect on the total mobility
only if the grain size is approximately of the same order as the mean free path of the charge carriers ($D \sim \lambda$). The mean free path
for the degenerate samples can be calculated from known mobility ($\mu$) and carrier concentration (n) using the following expression [131, 133]:
\begin{equation}\label{13}
  \lambda = (3\pi^{2})^{\frac{1}{3}}(\frac{\hbar\mu}{e})n^{\frac{1}{3}},
\end{equation}
The mean free path value calculated for the pure $SnO_{2}$ film is 1.257 nm which is considerable shorter than crystallite size (D $\sim$ 27 nm) calculated using XRD data. Moreover, the effect of crystallite interfaces is weaker in semiconductors, with n $\geq$ $10^{20}$ $cm^{-3}$, observed here, as a
consequence of the narrower depletion layer width at the interface between two grains [134]. Based on above discussion it is concluded that grain boundary scattering is not a dominant mechanism.

The mobility of the free carrier is not affected by surface scattering unless the mean free path is comparable to the film thickness [135]. Mean free path value calculated for the pure $SnO_{2}$ film is 1.257 nm, which is much smaller than the film thickness ($\sim$ 510 nm). Hence, surface scattering can be ruled out as the primary mechanism. Scattering by acoustical phonons [136] apparently plays a subordinate role in the pure $SnO_{2}$ films because no remarkable temperature dependence have been observed between 30 and $200^{o}$C. Moreover, neutral impurity scattering can be neglected because the neutral defect concentration is negligible in the pure $SnO_{2}$ films [131, 133]. Electron-electron scattering, as suggested to be important in Ref. 133, can also be neglected as it does not change the total electron momentum and thus not the mobility. In high crystalline $SnO_{2}$ films, scattering by dislocations and precipitation is expected to be of little importance [137].

Another scattering mechanism popular in unintentionally doped semiconductors is the ionized impurity scattering. According to the Brooks-Herring formula [138], the relaxation time for coupling to ionized impurities is in the degenerate case, given by
\begin{equation}\label{14}
  \tau_{i} = \frac{(2m^{*})^{\frac{1}{2}}(\epsilon_{o}\epsilon_{r})^{2}(E_{F})^\frac{3}{2}}{\pi e^{4}N_{i}f(x)},
\end{equation}
with $N_{i}$ the carrier concentration of ionized impurities and f(x) given by
\begin{equation}\label{15}
  f(x) = ln(1+x) - \frac{x}{1+x},
\end{equation}
with
\begin{equation}\label{16}
  x = \frac{8m^{*}E_{F}R_{S}^{2}}{\hbar^{2}},
\end{equation}
The screening radius $R_{S}$ is given by
\begin{equation}\label{17}
  R_{S} = (\frac{\hbar}{2e})(\frac{\epsilon_{o}\epsilon_{r}}{m^{*}})^{\frac{1}{2}}(\frac{\pi}{3N_{i}})^{\frac{1}{6}},
\end{equation}
where $\epsilon_{r}$ is the relative dielectric permittivity and $m^{*}$ is the effective mass of the carriers.
The mobility ($\mu$) is defined as
\begin{equation}\label{18}
\mu = \frac{e\tau}{m^{*}},
\end{equation}
Substitution of the $\tau_{i}$ expression [Eq. (14)] in Eq. (18) yields the expression for mobility due to ionized impurities as
\begin{equation}\label{19}
  \mu_{i} = \frac{(\frac{2}{m^{*}})^{\frac{1}{2}}(\epsilon_{o}\epsilon_{r})^{2}(E_{F})^{\frac{3}{2}}}{\pi e^{3}N_{i}f(x)},
\end{equation}
Since all the $H_{O}^{+}$ defects present in the pure $SnO_{2}$ films will be fully ionized at room temperature, impurity ion concentration will be equal to the free carrier concentration. Thus taking $N_{i}$ = n, $m^{*}$ = 0.31m, $\epsilon_{r}$ = 13.5 [73] and using Eq. (12) in Eq. (19) we get simplified form as
\begin{equation}\label{20}
  \mu _{i} = \frac{2.4232 \times 10^{-4}}{f(x)},
\end{equation}
with
\begin{equation}\label{21}
  x = 1.7942 \times 10^{-9} n^{\frac{1}{3}},
\end{equation}
The calculated mobility and measured mobility values for
pure $SnO_{2}$ thin films are 10.346 and 9.497 $cm^{2}V^{-1}s^{-1}$ respectively, both are comparable to each other. This clearly indicates that the main scattering mechanism reducing the intra-grain mobility of the electrons in pure $SnO_{2}$ films is the ionized impurity scattering. Ionized impurity scattering with singly ionized $H_{O}^{+}$ donors  best describes the mobility of pure $SnO_{2}$ samples. This finding supports our assumption that $H_{O}^{+}$ defect is source of conductivity in pure $SnO_{2}$ sample.

A Coulomb interactions between the electrons and the ionized impurities cause collisions and the scattering of the electrons. The amount of deflection depends on the speed of the electron and its proximity to the ion. The electrons will not experience the Coulombic force if they do not spend much time near the ionized impurities. If the electrons are moving fast, the amount of time spent in the location of the ionized impurity will be very short. This will affect the mobility of the electrons. Temperature of the material greatly affects ionized impurity scattering. At low temperatures, the electrons are moving slowly because thermal velocity is low. If the thermal velocity of the electrons is small, the amount of time the electrons stay within the vicinity of an ionized impurity increases. This increases the cross section of scattering. Another factor affecting ionized impurity scattering is the concentration of ionized impurities. The more ionized impurities in the material; the more scattering events occur. With the higher probability of scattering event, the mean time between collisions decrease, which decreases mobility. The mobility of electrons in non-degenerate semiconductor increases with temperature and is independent of the electron concentration, whereas the mobility in a highly degenerate semiconductor is nearly independent of temperature and increases with electron concentration [139, 140].
\begin{figure*}
  \centering
  \includegraphics[height=12cm, width=16cm]{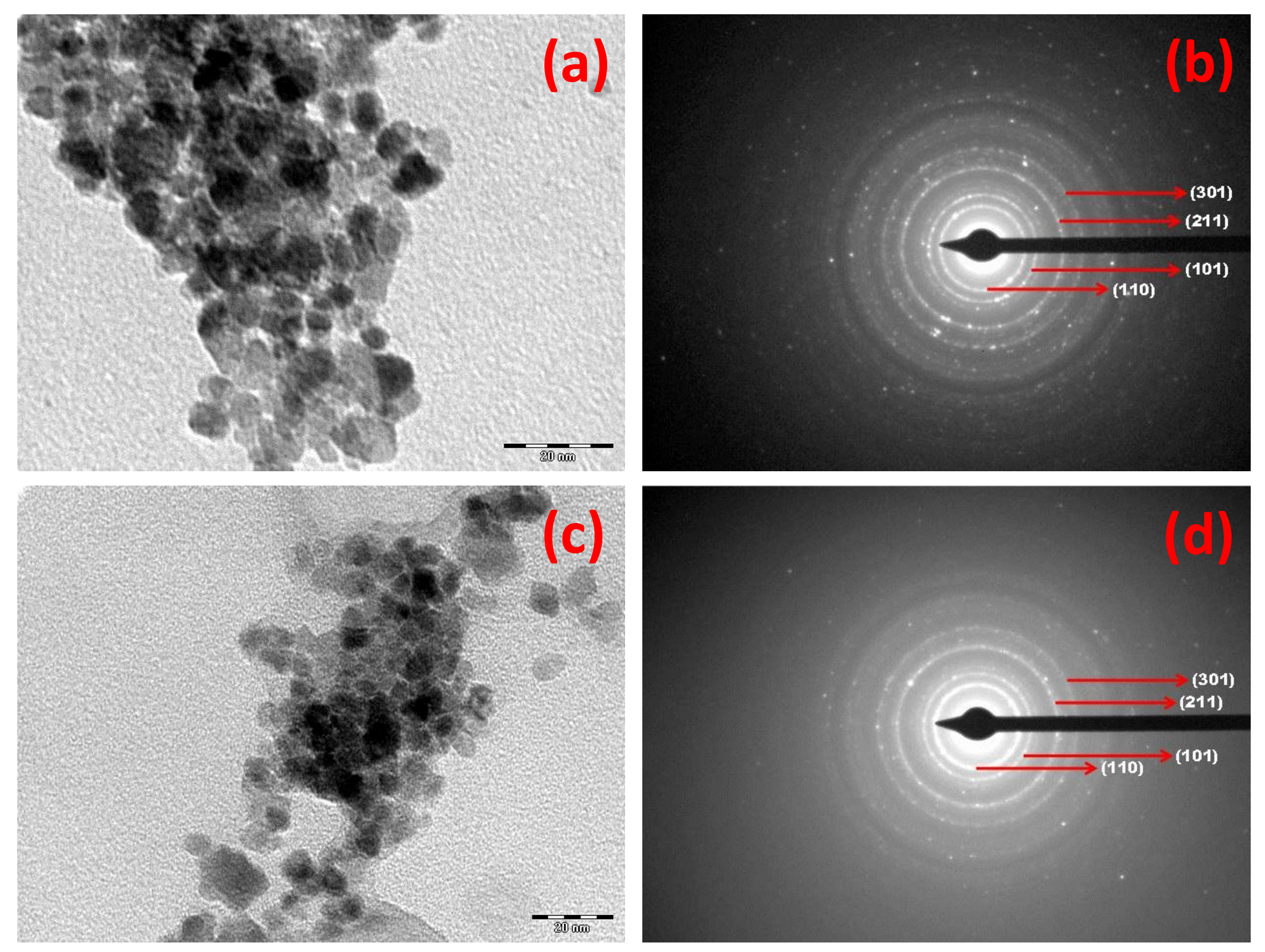}\\
  \caption{Transmission electron micrographs [(a) and (c)] of the pure $SnO_{2}$  and pristine $Sn_{0.9}Mn_{0.1}O_{2}$ thin films, respectively. Corresponding selected area electron diffraction (SAED) patterns for the pure $SnO_{2}$ and pristine $Sn_{0.9}Mn_{0.1}O_{2}$ thin films are shown in (b) and (d), respectively. Transmission electron micrographs of both thin films showing several nanocubes or nanospheres.}\label{10}
\end{figure*}
\begin{figure*}
  \includegraphics[height=8.3cm, width=16cm]{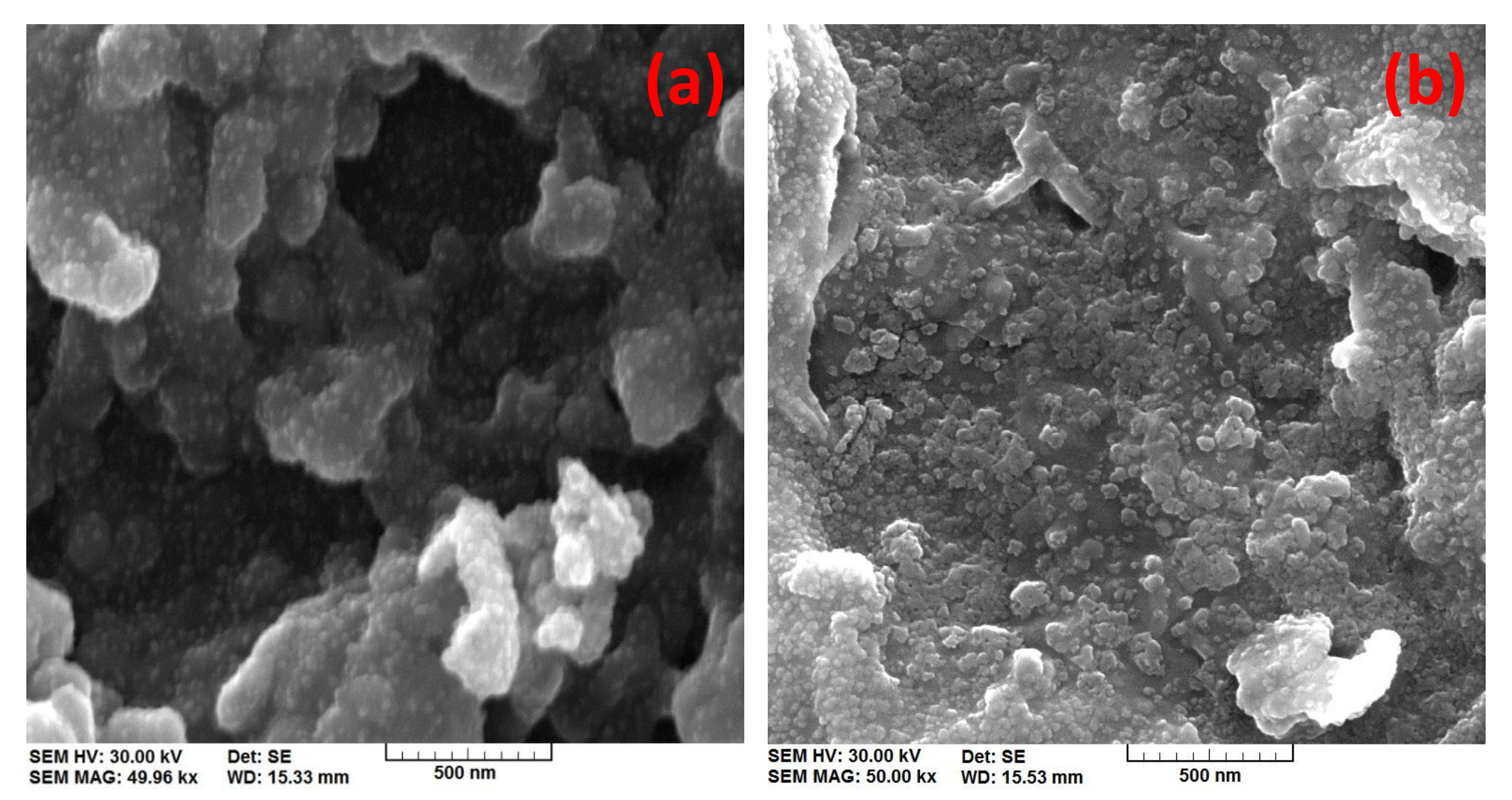}\\
  \caption{Scanning electron micrographs of the (a) pure $SnO_{2}$ and (b) pristine $Sn_{0.9}Mn_{0.1}O_{2}$ thin films.}\label{11}
\end{figure*}
\subsection{Microstructural properties}
\begin{figure*}
  \centering
  \includegraphics[height=7.79cm, width=16cm]{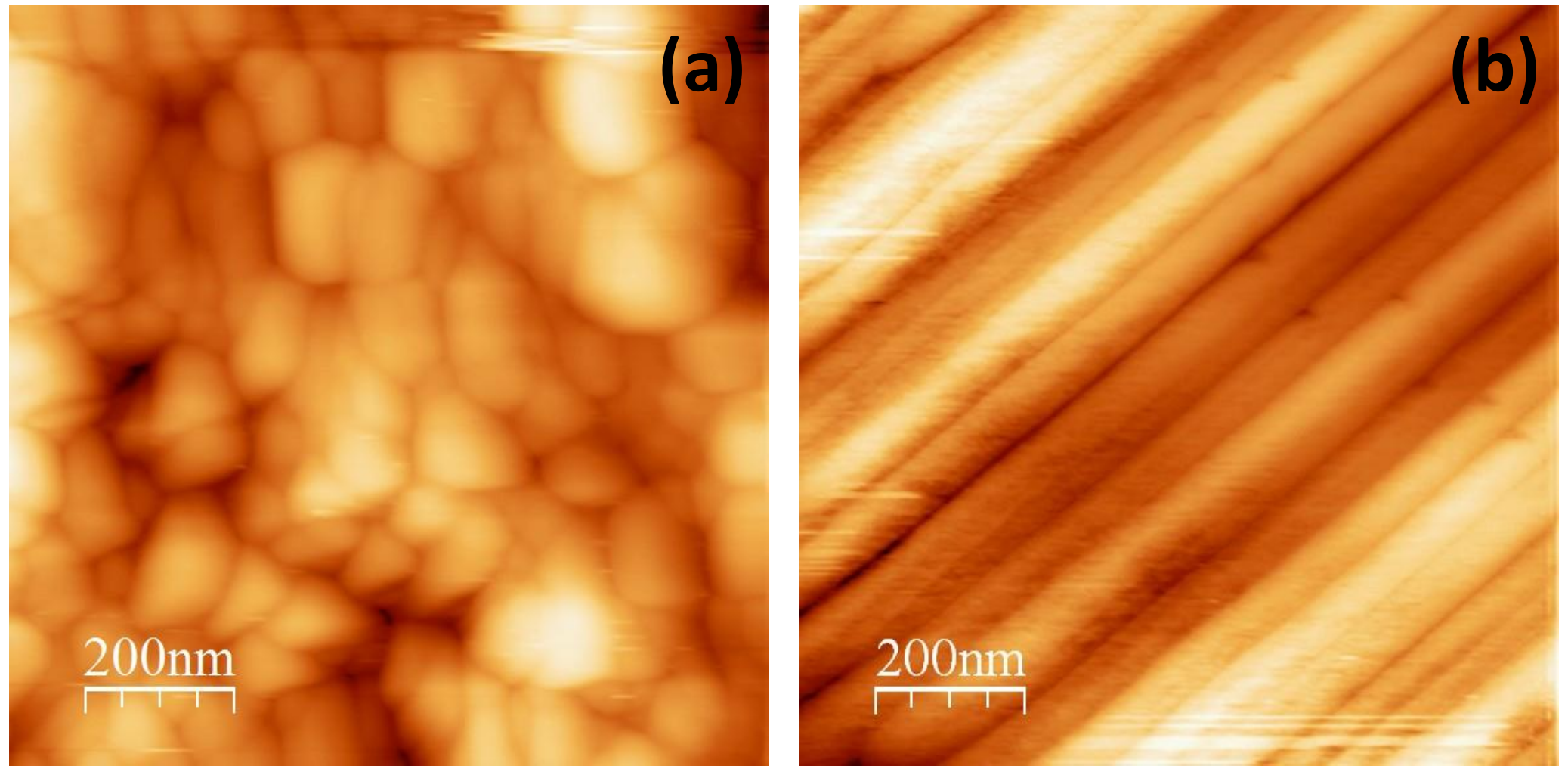}\\
  \caption{Atomic force microscopy images of $Sn_{0.9}Mn_{0.1}O_{2}$ surfaces: (a) annealed at $850^{o}$C after irradiation with $3\times10^{13}$ ions/$cm^{2}$, (b) irradiated with $3\times10^{13}$ ions/$cm^{2}$ after annealing at $850^{o}$C.}\label{12}
\end{figure*}
Structural characterization through TEM is a direct way of visualizing the estimated grain size. It also gives authentic information regarding the grain size distribution, crystalline nature, and other structural information. The transmission electron micrographs and the corresponding selected area electron diffraction (SAED) patterns for $SnO_{2}$ and $Sn_{0.9}Mn_{0.1}O_{2}$ samples are shown in Fig. 10. These TEM micrographs and SAED patterns have been analyzed using the IMAGE-J software. The TEM images of the $SnO_{2}$ and $Sn_{0.9}Mn_{0.1}O_{2}$ samples show the presence of interconnected nono-sized spheroidal grains. The crystallite size observed by TEM ($\sim$ 25 nm for $SnO_{2}$ and $\sim$ 30 nm for $Sn_{0.9}Mn_{0.1}O_{2}$) is in good agreement with that estimated by x-ray line broadening ($\sim$ 27 nm for $SnO_{2}$ and $\sim$ 29 nm for $Sn_{0.9}Mn_{0.1}O_{2}$). The SAED patterns shown in Figs. 10(b) \& 10(d) taken from $SnO_{2}$ and $Sn_{0.9}Mn_{0.1}O_{2}$ samples show several sharp rings, which are indexed to the (110), (101), (211) and (301) planes of the rutile crystalline structure of $SnO_{2}$. The electron diffraction pattern has been examined carefully for rings and spots of secondary phases, and it has been found that all the rings and spots belong to the tetragonal rutile structure of $SnO_{2}$ only. We have observed that there is no formation of any structural core-shell system.

The SEM images of $SnO_{2}$ and $Sn_{0.9}Mn_{0.1}O_{2}$ thin films deposited by chemical spray pyrolysis technique at substrate temperature of $450^{o}$C are shown in Fig. 11.  These SEM images reveal significant aggregation of the nano-grains. The small grains coagulated to form big clusters that reflect like the pomegranate structure [141]. This can be explained by the fact that smaller primary grains have a large surface free energy and would, therefore, tend to agglomerate faster and grow into larger particles/clusters. The aggregation of nano-grains is very important for the ferromagnetism in nano-crystalline dilute magnetic semiconductors [142], which can generate lattice defects and increase the domain size. In our samples, the aggregation should be promising for ferromagnetism. The aggregation makes it difficult to determinate the grain size accurately. The size that is estimated from the few spherical grains is about 25 nm. This is slightly smaller than that obtained from XRD measurements.

Fig. 12(a) shows the AFM image of $Sn_{0.9}Mn_{0.1}O_{2}$ film irradiated with $3\times10^{13}$ $ions/cm^{2}$ fluence and then annealed at $850^{o}C$ for 4 hr. The grains of this film are nearly spherical in shape with size in the range of 100 to 150 nm. The root mean square (rms) roughness is 16.5 nm. Fig. 12(b) shows the AFM image of the $Sn_{0.9}Mn_{0.1}O_{2}$ film annealed at $850^{o}C$ and then irradiated with $3\times10^{13}$ $ions/cm^{2}$ fluence. Morphology of this film is very much different from that shown in Fig. 12(a). At this condition nanorod/nanoribbon like structures with size in the range of 50-100 nm and rms roughness of 6.4 nm are grown in lateral direction on the surface. Morphology of irradiated film depends on the conductivity of target material [143]. If the target material is a conductor ($n \geq 10^{20}cm^{-3}$ for as-deposited $Sn_{0.9}Mn_{0.1}O_{2}$ film) then conduction electrons rapidly spread the energy of incident ion throughout the material and we find similar pristine/target sample like morphology with some dimensional change after irradiation. But if target material is an insulator (such as $Sn_{0.9}Mn_{0.1}O_{2}$ sample annealed at $850^{o}$C) then projectile ions can not spread their energy throughout the target material and therefore they create high energy region in the close vicinity of their paths. This cylindrical region around the path of the ion may become fluid if maximum temperature reached at the centre is higher than the melting temperature of the material. The structure of grains is changed (spherical to ribbon/rod like) only when melting is observed and the nature/size of the new structure corresponded closely to the maximum molten region. The typical amorphous nanoribbon/nanorod like structures may arise on the surface due to rapid resolidification of molten material [143]. Melt recoil pressure, surface tension, diffusion and evaporation dynamics may also effect the resolidification process of molten material and consequently the structure of grains.
\subsection{Band-gap studies}
\begin{figure}
  \centering
  \includegraphics[height=6.1cm, width=9.5cm]{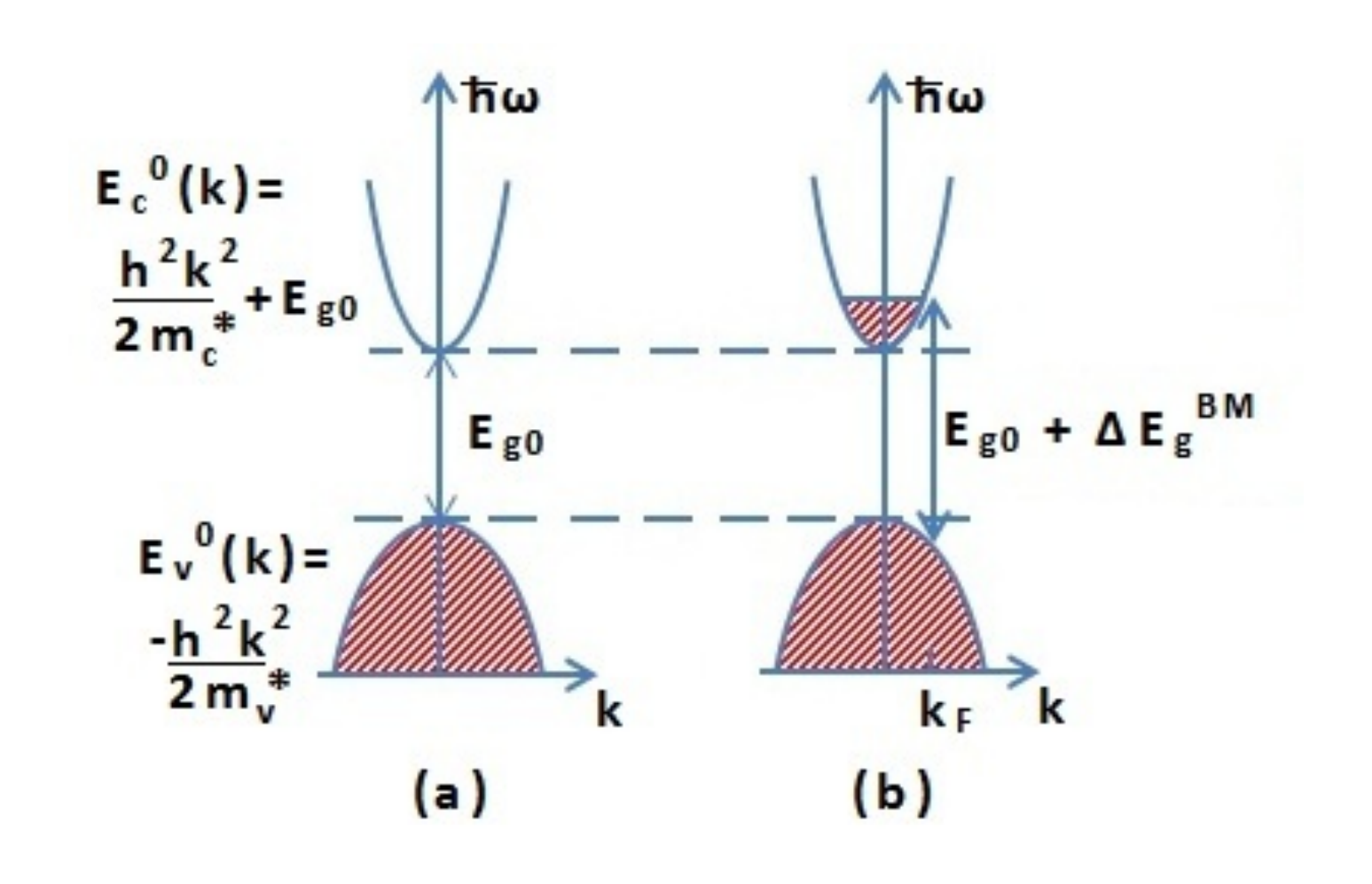}\\
  \caption{(a) Schematic band structure of stoichiometric $SnO_{2}$ showing the parabolic conduction and valence bands separated by the band gap $E_{g0}$ and (b) the Burstein-Moss shift $\Delta E_{g}^{BM}$ due to filling up of the lowest states in the conduction band. Shaded areas denote occupied regions. Band gaps, Fermi wave number, and dispersion relations are indicated.}\label{13}
\end{figure}
The presumed band structure of $SnO_{2}$ is shown in Fig. 13 with the parabolic conduction and valence bands. With the top of the valence band as reference energy, the dispersion relations for the unperturbed valence and conduction bands are
\begin{equation}\label{22}
  E_{v}^{0}(k) = \frac{-\hslash^{2}k^{2}}{2m_{v}^{*}},
\end{equation}
and
\begin{equation}\label{23}
  E_{c}^{0}(k) = E_{g0} + \frac{\hslash^{2}k^{2}}{2m_{c}^{*}},
\end{equation}
respectively. $E_{g0}$ is the intrinsic band gap of the semiconductor, k = $\frac{2\pi}{\lambda}$ is the wave number, and superscript 0 denotes unperturbed bands. When the carrier electron density n exceeds the Mott critical density [144], $n_{c}^{1/3} a_{0}^{*} \sim 0.25$ ($a_{0}^{*}$ is the effective Bohr radius), the low-energy levels of conduction band are filled up by the conduction electrons [145, 146]. Then photons with energy greater than the intrinsic band gap only get absorbed. Hence, the energy gap for direct transitions in the degenerate semiconductor is then given in terms of the unperturbed bands as
\begin{equation}\label{24}
  E_{g} = E_{c}^{0}(k_{F})-E_{v}^{0}(k_{F}),
\end{equation}
where $k_{F} = (3\pi^{2}n)^{\frac{1}{3}}$ is the Fermi wave number. Alternatively, we may write
\begin{equation}\label{25}
  E_{g} = E_{g0} + \Delta E_{g}^{BM},
\end{equation}
where the Burstein-Moss (BM) shift [145, 146] is given by
\begin{center}
$\Delta E_{g}^{BM} = [E_{c}^{0}(k_{F})-E_{v}^{0}(k_{F})] - E_{g0}$,
\end{center}
\begin{equation}\label{26}
  \Delta E_{g}^{BM} = \frac{\hslash^{2}}{2m_{vc}^{*}}(3\pi^{2}n)^{2/3},
\end{equation}
with the reduced effective mass
\begin{equation}\label{27}
  \frac{1}{m_{vc}^{*}} = \frac{1}{m_{v}^{*}} + \frac{1}{m_{c}^{*}},
\end{equation}
Neglecting the electron-electron and electron-defect interactions [147, 148], the band gap is
\begin{equation}\label{28}
  E_{g} = E_{g0} + \frac{\hslash^{2}}{2m_{vc}^{*}}(3\pi^{2}n)^{2/3}
\end{equation}
\begin{figure}
  \centering
  \includegraphics[height=6.93cm, width=9cm]{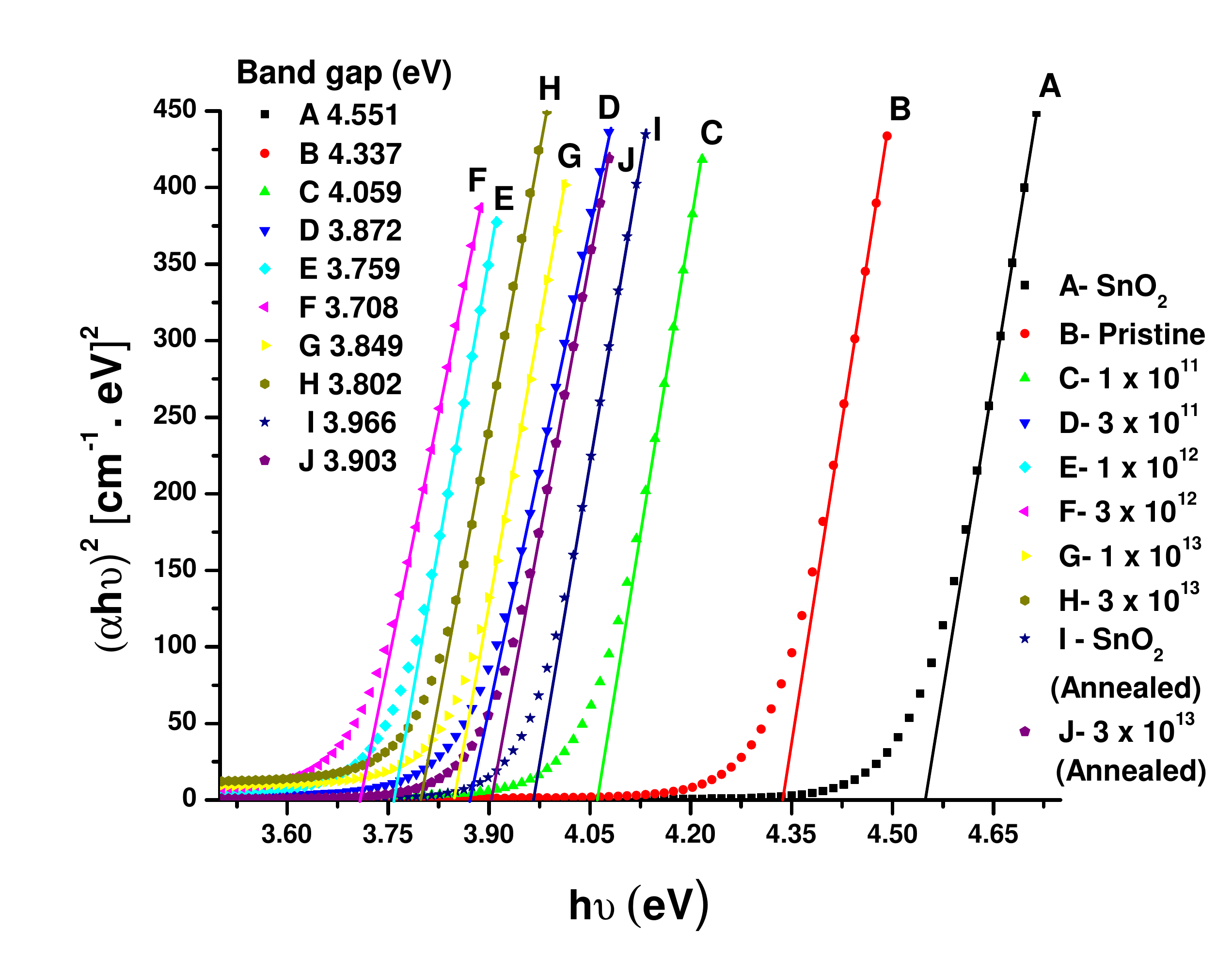}\\
  \caption{$(\alpha h\nu)^{2}$ vs $h\nu$ plots for the un-irradiated, irradiated and annealed samples. The direct energy band gap $E_{g}$ is obtained from the extrapolation to $\alpha$ = 0.}\label{14}
\end{figure}

To determine the absorption band-edge of $SnO_{2}$ films, we use the theory [149] developed for optical transitions in semiconductor, in which the absorption coefficient ($\alpha$) of $SnO_{2}$ is a parabolic function of the incident energy $h\nu$ and the optical band gap $E_{g}$,
\begin{equation}\label{29}
  \alpha (h\nu) = A(h\nu - E_{g})^{1/2}
\end{equation}
where A is a function of the refractive index of the material, the reduced mass, and the speed of light in vacuum. Using this relationship, the band edge of $SnO_{2}$ is evaluated in the standard manner from a plot of $(\alpha h\nu)^{2}$ versus photon energy $(h\nu)$. The extrapolation of the linear portion of the $(\alpha h\nu)^{2}$ vs. $h\nu$ plot to $\alpha$ = 0 gives the band gap value of the films. Fig. 14 shows such plots for all the samples and the linear fits obtained for these plots are also depicted in the same figure.
\begin{figure}
  \centering
  \includegraphics[height=6.93cm, width=9cm]{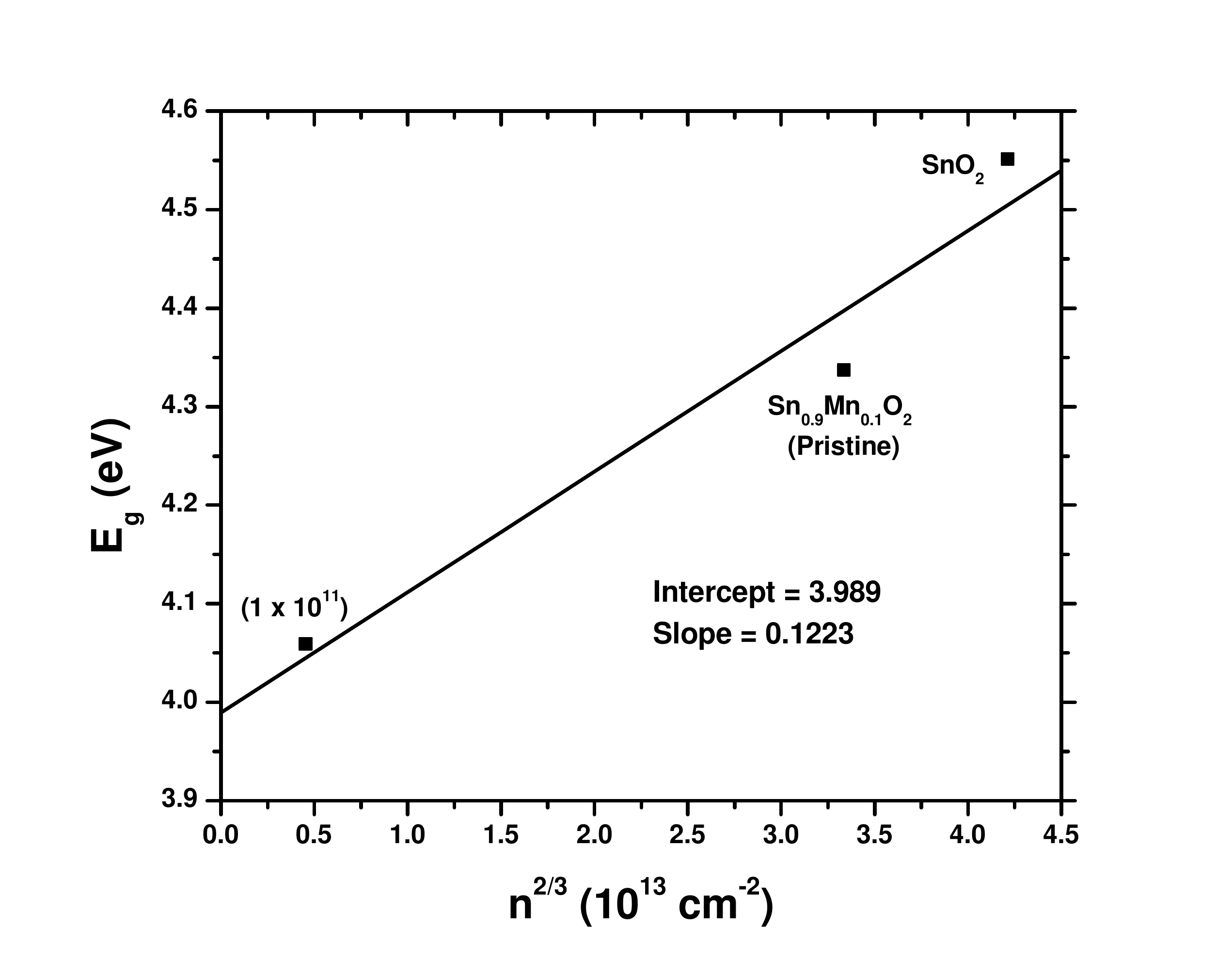}\\
  \caption{Band gap as a function of carrier concentration for the degenerate samples ($SnO_{2}$, $Sn_{0.9}Mn_{0.1}O_{2}$ and SHI: $1\times10^{11}$ $ions/cm^{2}$). The linear-fit of the experimental data gives $E_{g0}$ = 3.99 eV and $m^{*}_{vc}$ = 0.298m.}\label{15}
\end{figure}

The Burstein-Moss effect [145, 146], i.e., the widening of the band gap with increasing n, is exhibited when the free electron density (n) far exceeds the Mott critical density ($n_{c}$), whose magnitude can be estimated by Mott's criterion [144].
\begin{equation}\label{30}
  n_{c}^{1/3} a^{*}_{0} \sim 0.25
\end{equation}
The effective Bohr radius $a^{*}_{0}$ is given by:
\begin{equation}\label{31}
  a^{*}_{0} = \frac{h^{2}\epsilon _{0}\epsilon_{r}}{\pi e^{2}m^{*}_{c}}
\end{equation}
where $\epsilon_{r}$ is the static dielectric constant of the host lattice (equals 13.5 for $SnO_{2}$ [73, 148]), and $m^{*}_{c}$ is the effective mass of the electrons in the conduction band ($m^{*}_{c}$ = 0.31m [148]). Using these numbers, one can obtain $a^{*}_{0} \sim 2.31$ nm, and the critical
density $n_{c}$ is calculated as $1.3\times10^{18}$ $cm^{-3}$. Above this Mott critical density, the material is said to be degenerate.

For degenerate semiconductor (such as pure $SnO_{2}$, $Sn_{0.9}Mn_{0.1}O_{2}$ and $1\times10^{11}$ $ions/cm^{2}$), a plot of $E_{g}$ versus $n^{2/3}$ gives $E_{g0}$ as well as $m^{*}_{vc}$. Such a plot is given in Fig. 15. From this plot, the intrinsic band gap $E_{g0}$ and reduced effective mass $m^{*}_{vc}$ come out to be 3.99 eV and 0.298m, respectively. As the valence-band effective mass is given by
\begin{equation}\label{32}
  m^{*}_{v} = \frac{m^{*}_{c}.m^{*}_{vc}}{m^{*}_{c} - m^{*}_{vc}}
\end{equation}
Taking the conduction-band effective mass as $m^{*}_{c}$ = 0.31m, $m^{*}_{v}$ works out to be 7.69m for the present study. It may be pointed out that the reduced effective mass of $SnO_{2}$ obtained in the present work ($m^{*}_{vc}$ = 0.298m) is comparable than that estimated for $SnO_{2}$ by Sanon et. al. [148] ($m^{*}_{vc}$ = 0.237m) using $m^{*}_{c}$ = 0.31m and $m^{*}_{v}$ = 1.0m. A positive small value of valence-band effective mass ($m^{*}_{v}$ = 1.0m) can be obtained by including the many-body interactions in the valence-band and conduction bands. The band-gap narrowing due to electron-electron and electron-defect scattering in $SnO_{2}$ films is quite significant and cannot be neglected. The narrowing when added to experimental widening yields the actual Burstein-Moss widening and a positive small value of valence-band effective mass $m^{*}_{v}$ = 1.0m is thereby obtained.

The band gap obtained for annealed $SnO_{2}$ sample can be assumed as exact intrinsic band gap. Because the samples which have been annealed at a temperature of $850^{o}$C, they may be free of defects. On comparing the band gap of annealed $SnO_{2}$ sample ($\sim$ 3.97 eV) with that estimated from Fig. 15 ($\sim$ 3.99 eV), we find that both are nearly same. For degenerate samples such as $SnO_{2}$, $Sn_{0.9}Mn_{0.1}O_{2}$ and $1\times10^{11}$ $ions/cm^{2}$, it can be seen that the band gap is changing according to Burstein-Moss effect whereas for other irradiated samples (SHI: $3\times10^{11}$ to $3\times10^{13}$ $ions/cm^{2}$) we can notice a significant but small shift in band gap from intrinsic band gap ($E_{g0}$ = 3.97 eV). This noticeable shift in the band gap upon irradiation (SHI: $3\times10^{11}$ to $3\times10^{13}$ $ions/cm^{2}$) may be due to creation of intermediate energy levels [150-153].
\subsection{Magnetic properties}
For proper investigation and understanding of the magnetic properties of pristine and irradiated $Sn_{0.9}Mn_{0.1}O_{2}$ films, measurements of magnetization as a function of temperature [M(T)] and magnetic field [M(H)] were carried out over a temperature range of 5-300 K and field range of 0 to $\pm$ 2 T using a SQUID magnetometer. The magnetic field was applied parallel to the film plane. Figs. 16 and 17 show the magnetization versus applied magnetic field (M-H) curves measured at 5 and 300 K for the pristine and irradiated (SHI: $1\times10^{11}$ and $3\times10^{11}$ $ions/cm^{2}$) $Sn_{0.9}Mn_{0.1}O_{2}$ films with n = $1.9 \times 10^{20}$, $9.7 \times 10^{18}$ and $3.3 \times 10^{18}$ electrons $cm^{-3}$, respectively. The inset in Figs. 16 and 17 shows a zoom of the region of low magnetic fields that evidences the presence of a hysteresis. The saturation magnetization $(M_{S})$ is estimated to be $14.503\times10^{-2}$, $10.013\times10^{-2}$ and $3.745\times10^{-2}$ emu/g at 5 K and $13.793\times10^{-2}$, $8.265\times10^{-2}$ and $1.843\times10^{-2}$ emu/g at 300 K and the coercivity $(H_{C})$ is 89, 45 and 29 Oe at 5 K and  83, 35 and 24 Oe at 300 K for the pristine and irradiated (SHI: $1\times10^{11}$ and $3\times10^{11}$ $ions/cm^{2}$) $Sn_{0.9}Mn_{0.1}O_{2}$ samples by the M-H curves, respectively. The samples with higher carrier concentration (Pristine: $Sn_{0.9}Mn_{0.1}O_{2}$) show ferromagnetic characteristics with higher saturation magnetization and coercivity. On the other hand, thin films irradiated with higher fluence ($\geq 1\times10^{12}$ $ions/cm^{2}$) are diamagnetic at 5 K, as shown in the inset of Fig. 18.
\begin{figure}
  \centering
  \includegraphics[height=7.3cm, width=9.5cm]{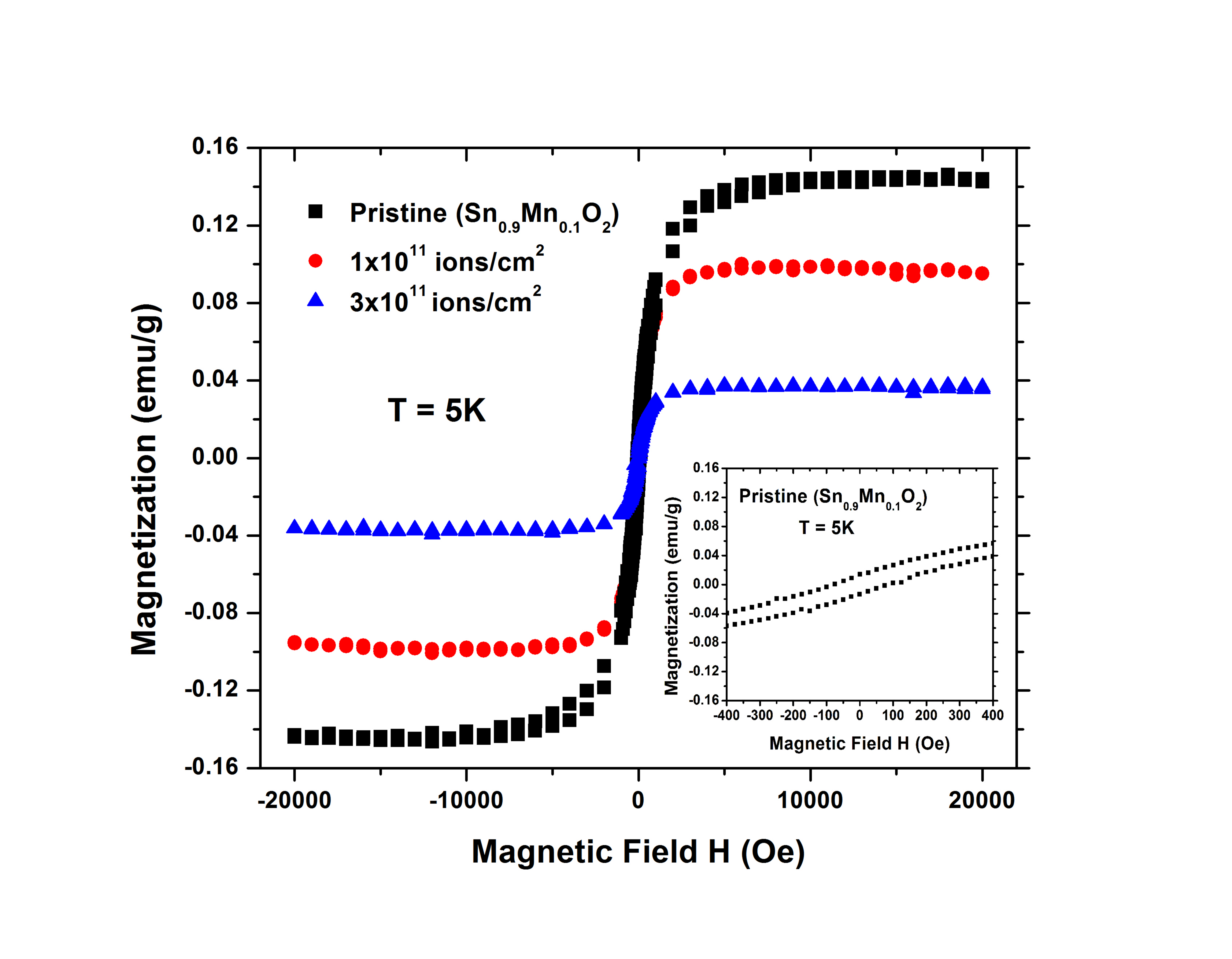}\\
  \caption{Field-dependent magnetization of the pristine and irradiated (SHI: $1\times10^{11}$ and $3\times10^{11}$ $ions/cm^{2}$) $Sn_{0.9}Mn_{0.1}O_{2}$ thin films measured at 5K. The inset shows the low-field part in an enlarged scale that evidences the presence of a hysteresis in pristine $Sn_{0.9}Mn_{0.1}O_{2}$ sample.}\label{17}
\end{figure}
\begin{figure}
  \centering
  \includegraphics[height=7.3cm, width=9.5cm]{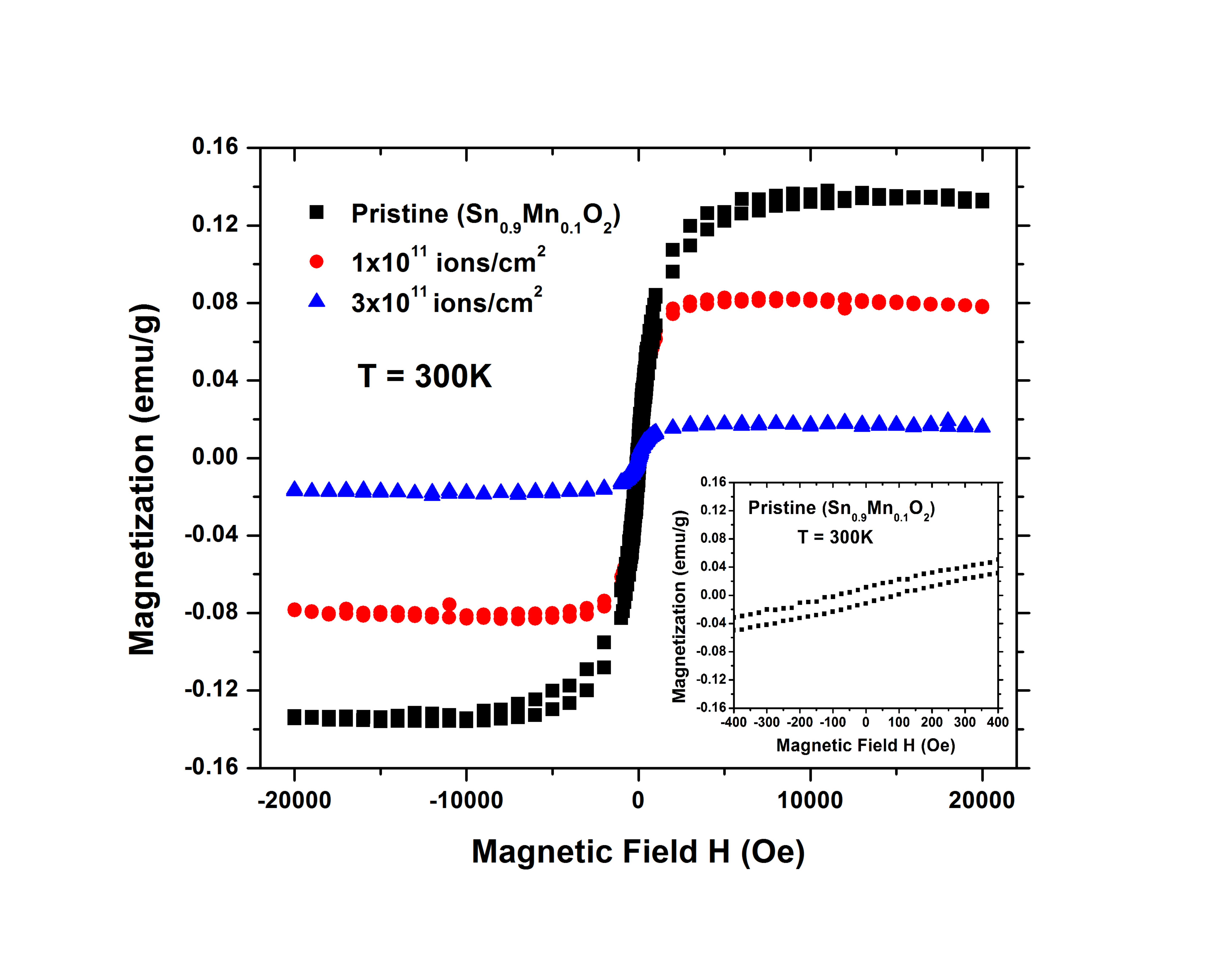}\\
  \caption{Field-dependent magnetization of the pristine and irradiated (SHI: $1\times10^{11}$ and $3\times10^{11}$ $ions/cm^{2}$) $Sn_{0.9}Mn_{0.1}O_{2}$ thin films measured at 300K. The inset shows the low-field part in an enlarged scale that evidences the presence of a hysteresis in pristine $Sn_{0.9}Mn_{0.1}O_{2}$ sample.}\label{18}
\end{figure}
\begin{figure}
  \centering
  \includegraphics[height=7.3cm, width=9.5cm]{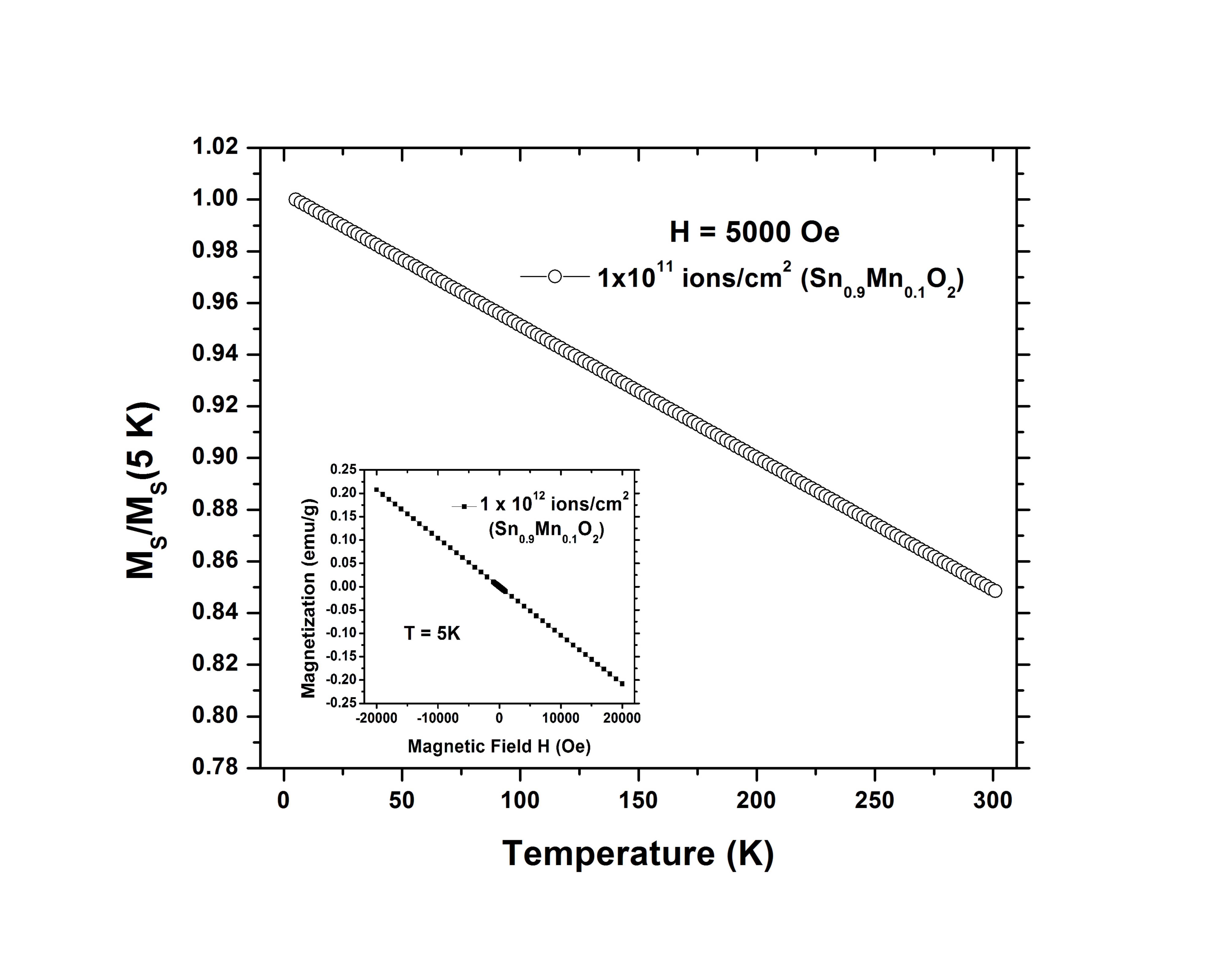}\\
  \caption{Normalized $M_{S}(T)$ plot with H = 5000 Oe for $Sn_{0.9}Mn_{0.1}O_{2}$ sample irradiated at $1\times10^{11}$ $ions/cm^{2}$ fluence. The inset shows the 5K field-dependent magnetization of $Sn_{0.9}Mn_{0.1}O_{2}$ sample irradiated at $1\times10^{12}$ $ions/cm^{2}$ fluence.}\label{19}
\end{figure}
The diamagnetic films have $n \leq 7.5 \times 10^{16}$ electrons $cm^{-3}$, which is much lower than that observed in ferromagnetic pristine films. We have observed that the films are ferromagnetic for the higher and intermediate range of n, $1.9 \times 10^{20}$ - $3.3 \times 10^{18}$ electrons $cm^{-3}$, and diamagnetic for lower range, $n \leq 7.5 \times 10^{16}$ electrons $cm^{-3}$. Thus, lowering the n from $1.9 \times 10^{20}$ to $7.5 \times 10^{16}$ electrons $cm^{-3}$ results in both electrical and magnetic phase transitions, a transition from a ferromagnetic semiconducting state to a diamagnetic insulating state.
The diamagnetic background of the pure $SnO_{2}$ film and substrate has been subtracted from all of the magnetization data shown here. The correlation between the magnetic and electrical properties in the present $Sn_{0.9}Mn_{0.1}O_{2}$ system formulates it to be a potential aspirant for the spintronics oriented devices wherein the communication between the spin and charge is highly desired.

For the pristine film, the M-H curve at 5 K differed by less than 6\% from that measured at room temperature (300 K), from which we guess that the Curie temperature $(T_{C})$ of pristine sample is well above room temperature. But, in the case of $1\times10^{11}$ and $3\times10^{11}$ $ions/cm^{2}$ irradiated films, the M-H curves at 5 K differed by 16\% and 52\% from those measured at room temperature, respectively. This clearly indicates that the Curie temperature of the irradiated film is less as compared to pristine film. We also recorded the M vs T curves of these samples in a field of 0.5 T. Fig. 18 shows the normalized $M_{S}(T)$/$M_{S}(5K)$ data from 5-300 K, for a $Sn_{0.9}Mn_{0.1}O_{2}$ film irradiated at $1\times10^{11}$ $ions/cm^{2}$ fluence. The absence of any sharp drop in the normalized $M_{S}(T)$ plot suggests that the film is ferromagnetic with a Curie temperature exceeding 300 K.

The appearance of room temperature ferromagnetism in pristine and irradiated $Sn_{0.9}Mn_{0.1}O_{2}$ samples cannot be due to the presence of secondary phases, because the metallic manganese and nearly all of the possible manganese-based binary and ternary oxide phases (MnO, $MnO_{2}$ and $Mn_{2}O_{3}$) are antiferromagnetic with a Neel temperature that is much less than 300 K [21]. However, $SnMn_{2}O_{4}$ and $Mn_{3}O_{4}$ phases are exceptions; they are ferromagnetic with Curie temperatures of 46 K and 53 K, respectively [25, 29, 154-156]. In the present work, the electron and x-ray diffraction analyses have not revealed any manganese oxide phases, although x-ray diffraction technique is not sensitive enough to detect secondary phases, if present at a very minute level. Even if these ferromagnetic $SnMn_{2}O_{4}$ and $Mn_{3}O_{4}$ phases are present, these cannot responsible for the ferromagnetic behavior observed at room temperature in pristine and irradiated $Sn_{0.9}Mn_{0.1}O_{2}$ thin films.
\section{Conclusions}
High crystalline thin films of $SnO_{2}$ and $Sn_{0.9}Mn_{0.1}O_{2}$ have been deposited by the spray pyrolysis technique and then deposited  $Sn_{0.9}Mn_{0.1}O_{2}$ films have been irradiated with 120 MeV $Au^{9+}$ ions to study the modification of structural, microstructural, electrical and magnetic properties. The threshold value of electronic energy loss $(S_{eth})$ for $Sn_{0.9}Mn_{0.1}O_{2}$ has been calculated according to Szenes' thermal spike model. The electronic energy loss ($S_{e}$) of 120 MeV $Au^{9+}$ ions in $Sn_{0.9}Mn_{0.1}O_{2}$ is less than the threshold electronic energy loss $(S_{eth})$ required for the track formation in $Sn_{0.9}Mn_{0.1}O_{2}$ film. Therefore, we expect that only point defects or clusters of point defects will be produced in $Sn_{0.9}Mn_{0.1}O_{2}$ thin films after irradiation. Efforts have been made to calculate equilibrium substrate/film temperature using the Stefan's equation and the calculated temperature (834 K) has been found to be quite higher than the crystallization temperature of tin oxide. This equilibrium temperature is expected to be develop within the grains of $Sn_{0.9}Mn_{0.1}O_{2}$ during irradiation period and facilitate (i) increase of grain size, (ii) removal of micro-strain, and (iii) migration of point defects. The increase rate of the substrate/film temperature during irradiation has been determined by the difference between the input power density and the heat dissipation via radiation, divided by the heat capacity of the substrate. Through this it has been concluded that, in the case of high fluence, irradiation period is much longer than the amount of time needed to achieve equilibrium temperature. Therefore, in this case, the samples spend their almost full irradiation period at equilibrium temperature. This equilibrium temperature can influence the physical properties of high fluence irradiated samples. X-ray and electron diffraction patterns analysis reveals that all the unirradiated and irradiated thin films are pure crystalline with tetragonal rutile phase of tin oxide which belongs to the space group P$4_{2}$/mnm (number 136). The Williamson-Hall (W-H) method has been used to evaluate the crystallite size and the microstrain of all the samples. The average crystallite size of $SnO_{2}$/$Sn_{0.9}Mn_{0.1}O_{2}$ nanoparticles estimated from W-H analysis and TEM/SEM analysis is highly inter-correlated. Electrical measurement shows that as-deposited $SnO_{2}$ and $Sn_{0.9}Mn_{0.1}O_{2}$ films are in conducting state with n = $2.735 \times 10^{20}$ $cm^{-3}$ and $1.927 \times 10^{20}$ $cm^{-3}$, respectively. The results of electrical measurements suggest that $H_{O}^{+}$ defects are responsible for the conductivity in as-deposited thin films. Through electrical investigation it has also been found that the main scattering mechanism reducing the intra-grain mobility of the electrons in pure $SnO_{2}$ films is the ionized impurity scattering. Ionized impurity scattering with singly ionized $H_{O}^{+}$ donor best describes the mobility of pure $SnO_{2}$ samples. Measurement of resistivity, mobility, and carrier density as a function of ion fluence ($1\times10^{11}$ to $3\times10^{13}$ $ions/cm^{2}$) promulgates that increasing fluence results in degradation in electrical properties of $Sn_{0.9}Mn_{0.1}O_{2}$ film. Typical TEM micrographs of as-deposited $SnO_{2}$ and $Sn_{0.9}Mn_{0.1}O_{2}$ thin films show well isolated and highly crystallized spherical shaped crystallites. Electron diffraction patterns taken from several crystallites confirm the $SnO_{2}$ structure in both $SnO_{2}$ and $Sn_{0.9}Mn_{0.1}O_{2}$ samples and no evidence for secondary phases are observed. The SEM micrographs of as-deposited $SnO_{2}$ and $Sn_{0.9}Mn_{0.1}O_{2}$ samples reveal significant aggregation of the nano-particles. Our AFM study demonstrates that the morphologies of irradiated films are linked with carrier concentration of target materials. If the target material is a conductor ($n \geq 10^{20}cm^{-3}$ for as-deposited $Sn_{0.9}Mn_{0.1}O_{2}$ film) then conduction electrons rapidly spread the energy of incident ion throughout the material and we find similar pristine/target sample like morphology with some dimensional change after irradiation. But if target material is an insulator (such as $Sn_{0.9}Mn_{0.1}O_{2}$ sample annealed at $850^{o}$C) then projectile ions can not spread their energy throughout the target material and therefore they create high energy region in the close vicinity of their paths. This cylindrical region around the path of the ion may become fluid if maximum temperature reached at the centre is higher than the melting temperature of the material. The structure of grains is changed (spherical to rod like) only when melting is observed and the nature/size of the new structure corresponded closely to the maximum molten region. In the present case, amorphous nanoribbon/nanorod like structures may arise on the surface of the irradiated films due to rapid resolidification of the molten material. The optical band gap ($E_{g}$) of the films has been determined from the spectral dependence of the absorption coefficient $\alpha$ by the application of conventional extrapolation method (Tauc plot). For degenerate samples such as $SnO_{2}$, $Sn_{0.9}Mn_{0.1}O_{2}$ and $1\times10^{11}$ $ions/cm^{2}$, it can be seen that the band gap is changing according to Burstein-Moss effect whereas for other irradiated samples (SHI: $3\times10^{11}$ to $3\times10^{13}$ $ions/cm^{2}$) we can notice a significant but small shift in band gap from intrinsic band gap ($E_{g}$ = 3.97 eV). This noticeable shift in the band gap upon irradiation (SHI: $3\times10^{11}$ to $3\times10^{13}$ $ions/cm^{2}$) may be due to creation of intermediate energy levels. Intrinsic room temperature ferromagnetism has been observed in all degenerate films (such as pristine and irradiated (SHI: $1\times10^{11}$ and $3\times10^{11}$ $ions/cm^{2}$) $Sn_{0.9}Mn_{0.1}O_{2}$ thin films). The variation of $M_{S}$ values suggest that room temperature ferromagnetism could be enhanced by increasing carrier concentration of the films. No evidence of any impurity phases are detected in $Sn_{0.9}Mn_{0.1}O_{2}$ suggesting that the emerging ferromagnetism in this system is most likely related to the properties of host $SnO_{2}$ system.
\begin{acknowledgments}
The authors gratefully acknowledge to V. Ganesan, A. Banerjee, N. P. Lalla, M. Gupta, M. Gangrade, UGC-DAE Consortium for Scientific Research, Indore and D. Kanjilal, D. K. Avasthi, A. Tripathi, K. Asokan, P. K. Kulriya, I. Sulania, IUAC, New Delhi for providing the characterization facilities. We also thank Pelletron group, IUAC, New Delhi for expert assistance in the operation of the tandem accelerator.
\end{acknowledgments}

\end{document}